 \documentclass[final,5p,times,twocolumn]{elsarticle}

\usepackage{lineno}

\usepackage[final]{changes}

\usepackage{amsmath,amssymb,amsfonts}
\usepackage{algorithmic}
\usepackage{wrapfig}
\usepackage{graphicx}
\usepackage{textcomp}
\usepackage{xcolor}
\def\BibTeX{{\rm B\kern-.05em{\sc i\kern-.025em b}\kern-.08em
		T\kern-.1667em\lower.7ex\hbox{E}\kern-.125emX}}

\graphicspath{{../pdf/}{../jpeg/}{figures/}}

\newcommand{\PLH}{{\mkern-2mu\times\mkern-2mu}} 

\usepackage{algorithm}
\usepackage{comment}

\usepackage{makecell}

\usepackage{csquotes}

\usepackage[export]{adjustbox}

\usepackage{booktabs}   
\usepackage{subcaption} 
\usepackage{multirow}

\usepackage[hyphens]{url}

\makeatletter
\newcommand*{\textoverline}[1]{$\overline{\hbox{#1}}\m@th$}
\makeatother

\usepackage{makecell}
\usepackage{xcolor}
\usepackage[normalem]{ulem}

\newcommand\ore[1]{\noindent{\color{black} {#1}}}
\newcommand\reduline{\bgroup\markoverwith
	{\textcolor{magenta}{\rule[0.3ex]{2pt}{2.4pt}}}\ULon}

\journal{Computers and Electrical Engineering}

\makeatletter
\def\ps@pprintTitle{%
	\let\@oddhead\@empty
	\let\@evenhead\@empty
	\def\@oddfoot{\textcopyright~2020. This work is licensed under a CC BY-NC-ND 4.0 International License:\\ http://creativecommons.org/licenses/by-nc-nd/4.0/}%
	\let\@evenfoot\@oddfoot}
\makeatother

\begin{document}
	
		\begin{frontmatter}
		
		
		\title{Accelerating Sparse Matrix-Matrix Multiplication with GPU Tensor Cores}
		


		%
		
		\author[1]{Orestis~Zachariadis\corref{*}}
		\cortext[*]{Corresponding author}
		\ead{orestis.zachariadis@uco.es}
        \author[1]{Nitin Satpute}
		\author[2]{Juan~G\'omez-Luna}
		\author[1]{Joaqu\'in~Olivares}
		
		\address[1]{Department of Electronic and Computer Engineering, Universidad de Cordoba, C\'ordoba, Spain}
		\address[2]{Department of Computer Science, ETH Zurich, Zurich, Switzerland}


\begin{abstract}
	Sparse general matrix-matrix multiplication (spGEMM) is an essential component in many scientific and data analytics applications. However, the sparsity pattern of the input matrices and the interaction of their patterns make spGEMM challenging.
	Modern GPUs include Tensor Core Units (TCUs), which specialize in dense matrix multiplication. Our aim is to re-purpose TCUs for sparse matrices.
	%
	The key idea of our spGEMM algorithm, tSparse, is to multiply sparse rectangular blocks using the mixed precision mode of TCUs.
	tSparse partitions the input matrices into tiles and operates only on tiles which contain one or more elements. It creates a task list of the tiles, and performs matrix multiplication of these tiles using TCUs. To the best of our knowledge, this is the first time that TCUs are used in the context of spGEMM.
	We show that spGEMM, with our tiling approach, benefits from TCUs.
	Our approach significantly improves the performance of spGEMM in comparison to cuSPARSE, CUSP, RMerge2, Nsparse\replaced{, AC-SpGEMM and spECK}{}.
\end{abstract}
		
	\begin{keyword}
		
		Sparse Matrix Multiplication \sep GPU \sep Tensor Cores \sep Parallel Computing \sep SpGEMM
		
		
		
	\end{keyword}
		
	\end{frontmatter}
	

\section{Introduction}

Sparse general matrix-matrix multiplication (spGEMM), similar to its dense counterpart, performs the Matrix Multiplication (MM) of two sparse matrices. The main difference between sparse and dense MM is that we have to account for sparse matrices, which contain mostly zero elements. spGEMM is an important component in scientific and data analytics applications\replaced{ such as}{.} Graph analytics \cite{davis2018algorithm}, Bread\ore{th}-First-Search (BFS) \cite{gilbert2006high}, Algebraic Multigrid (AMG) \cite{bell_exposing_2012}, Schur complement \cite{yamazaki2010techniques}, etc.\replaced{}{ use sparse matrices.} Sparse matrices, which often \replaced{contain millions of elements}{contain million of elements}, require MM methods that do not waste computing resources on elements that are zero. The diverse structure and density of sparse matrices pose difficulties in regards to memory management and load balancing in parallel systems \cite{cusp_optimizing_2015, bhsparse_2015, winter_adaptive_2019, RMerge_2015, RMerge2_2018}.

The re-emergence of deep learning motivated the creation of application specific integrated circuits (ASIC) that specialize in MM. MM is a core component of convolution \cite{vasudevan2017parallel}\replaced{ and these}{.} ASICs accelerate the calculation of the convolutional layers significantly in comparison to the normal, general purpose, processing cores. Such ASICs are Tensor Core Units (TCUs) from NVIDIA \cite{CUDAProgrGuide} and TPUs from Google \cite{google_tpu}. We utilize\replaced{ the}{} TCUs\replaced{ from NVIDIA}{} to accelerate spGEMM\replaced{}{. TCUs from NVIDIA provide an attractive target} for two reasons. First, accessibility. They are widely available as they are included in the new generation of GPUs from NVIDIA. Second, mixed precision. Mixed precision allows TCUs to mix 16-bit inputs and 32-bit multiplication and accumulation. Typically, in regards to deep learning 16-bits of precision (or less) is sufficient for training purposes and therefore tensor unit manufacturers opt only for 16-bit or lower precisions. Therefore, mixed precision of TCUs widens the application field to scientific problems which are more demanding w.r.t. precision.

Our work is motivated by three key observations. First, blocking sparse matrix storage formats \cite{bmsparse_2018}, which group the elements of the matrix into rectangular \emph{tiles}, are a good fit for TCUs which expect rectangular matrices as input. Second, TCUs are very efficient even when they are not fully occupied \cite{dakkak2019accelerating}. Third, even though TCUs support only low precision inputs, they can operate in mixed precision mode to perform operations that require higher precision, e.g., GEMM \cite{haidar_harnessing_2018,markidis_nvidia_2018}. The key idea of our approach is to partition the input in tiles and multiply the tiles with TCUs. Tiles are sparse, but TCUs perform MM efficiently even when not fully occupied. Mixed precision mode is necessary in order to keep sufficient accuracy when multiplying large matrices.

Based on these observations, we propose a new GPU-based framework for spGEMM computation. Our novel methodology groups elements into tiles and uses the fast MM of TCUs to multiply the tiles. To the best of our knowledge, this is the first proposal of using TCUs in the context of spGEMM. \replaced{}{We name our TCU-based spGEMM \emph{tSparse}.}
Our \replaced{TCU-based spGEMM}{} methodology\replaced{, which we name \emph{tSparse},}{} has two advantages. First, it takes advantage of fast MM of TCUs. Second, by utilizing TCUs to process the MM, we leave the normal processing cores free for non-canonical workloads.

tSparse modifies Expand-Sort-Compress (ESC) methodology \cite{bell_exposing_2012} of CUSP \cite{Cusp} to \emph{Sort-Expand and Compress} (SEaC), i.e., tSparse brings both multiplication and accumulation after the sorting step. The benefits of this change are twofold. First, tSparse does not have to maintain in memory a large matrix for the intermediate products. Second, tSparse takes full advantage of Multiplication-Accumulation (MAC) operation of TCUs. To that end, we form a task list instead of calculating the intermediate products immediately. Our GPU kernels consume tiles from the task list and perform MM of the tiles using TCUs.

We measure the performance of tSparse in matrix squaring ($ A*A $) on matrices from SuiteSparse (formerly known as University of Florida Sparse Matrix Collection) \cite{suitesparse}.
We compare the performance of our approach to \ore{four} state-of-the-art libraries: cuSPARSE\ore{ \cite{cusparse_presentation_2012}}, CUSP \cite{Cusp}, RMerge2 \cite{RMerge2_2018}, Nsparse \cite{nagasaka_high-performance_2017}, AC-SpGEMM \cite{winter_adaptive_2019} and spECK \cite{parger_speck_2020}.


The rest of the paper is organized as follows. Section \ref{sec:background} gives a background on spGEMM. Section \ref{sec:overview} presents a high-level overview of tSparse, whereas Section \ref{sec:impl} describes tSparse in-depth. Section \ref{sec:eval} introduces our test configuration, which we use in Section \ref{sec:results} to evaluate the performance of tSparse. Section \ref{sec:related} presents related work. Finally, Section \ref{sec:conclusion} concludes the paper.

\section{Background} \label{sec:background}
In this section we describe: 1) the sparse matrix storage format we use (Section \ref{sec:format}\ore{)}, 2) the sparse matrix-matrix multiplication (Section \ref{sec:spgemm}\ore{)}, 3) the challenges in spGEMM, 4) the ESC methodology of CUSP (Section \ref{sec:esc}\ore{)}, 5) precision of real numbers (Section \ref{sec:reals}\ore{)}, and 6) \replaced{the tensor cores of Nvidia}{} (Section \ref{sec:tcus}).

\subsection{Storage format} \label{sec:format}
In sparse matrices, typically, the number of non-zero (NNZ) elements is much smaller than the number of zero elements. In order to save memory without degrading \replaced{the}{hte} performance of MM, we need an efficient way to store only non-zero (NZ) elements.

\subsubsection{COO format}
COO format stores each NZ element along with the coordinates the element would have in the dense representation of the matrix\replaced{}{. If the sparse matrix is sufficiently sparse, the storage cost of saving the coordinates is much smaller than the dense representation. This storage format is \replaced{}{formally} known as COO format (Coordinate Format), one of the simplest and most used storage formats} \cite{barrett1994templates}. \replaced{The COO format uses three arrays: for elements, for row indices and for column indices.}{The COO format uses three arrays: 1) for row indices, 2) for column indices, 3) for elements.}


\subsubsection{Bitmap format} \label{ssec:bitmap}
TCUs simultaneously process multiple elements in rectangular structures. However, COO stores only single elements and has no concept of rectangular structures, therefore it is not sufficient by itself as a storage format for tSparse. \ore{For this reason}, we use a bitmap-based block shaped storage format to store sparse matrices \cite{bmsparse_2018}.

In our work, we use a bitmap format similar to \cite{bmsparse_2018} for three reasons: 1) it is simple and straightforward, 2) square tiles of fixed size fit well to TCUs, and 3) the performance of the format has been evaluated in \cite{bmsparse_2018}. \ore{The basic idea is to partition the matrices in a grid of $ 8 \PLH 8 $ square blocks and work only on non-empty blocks. We refer to these blocks as \emph{tiles}. The elements inside each tile have the same placement as they have in the dense representation of the matrix. Each element in the tile can be either zero or NZ. To keep track which elements are NZ we use a bitmap, a binary number of which each digit corresponds to one slot of the tile. If a slot contains a NZ we set the respective bit of the bitmap to \enquote{1}, otherwise to \enquote{0}. Then, tiles are stored in COO format. The difference with the standard COO format is that, instead of using \ore{single} elements as values, now we use a tuple\ore{pair} of two values: 1) an \emph{index} to the \emph{element array}. The element array holds the elements of the tile (elements of the same tile are in consecutive positions of the array), and 2) the \emph{bitmap} of the tile.}

\begin{figure*}[h!]
	\centering
	\includegraphics[width=0.7\linewidth]{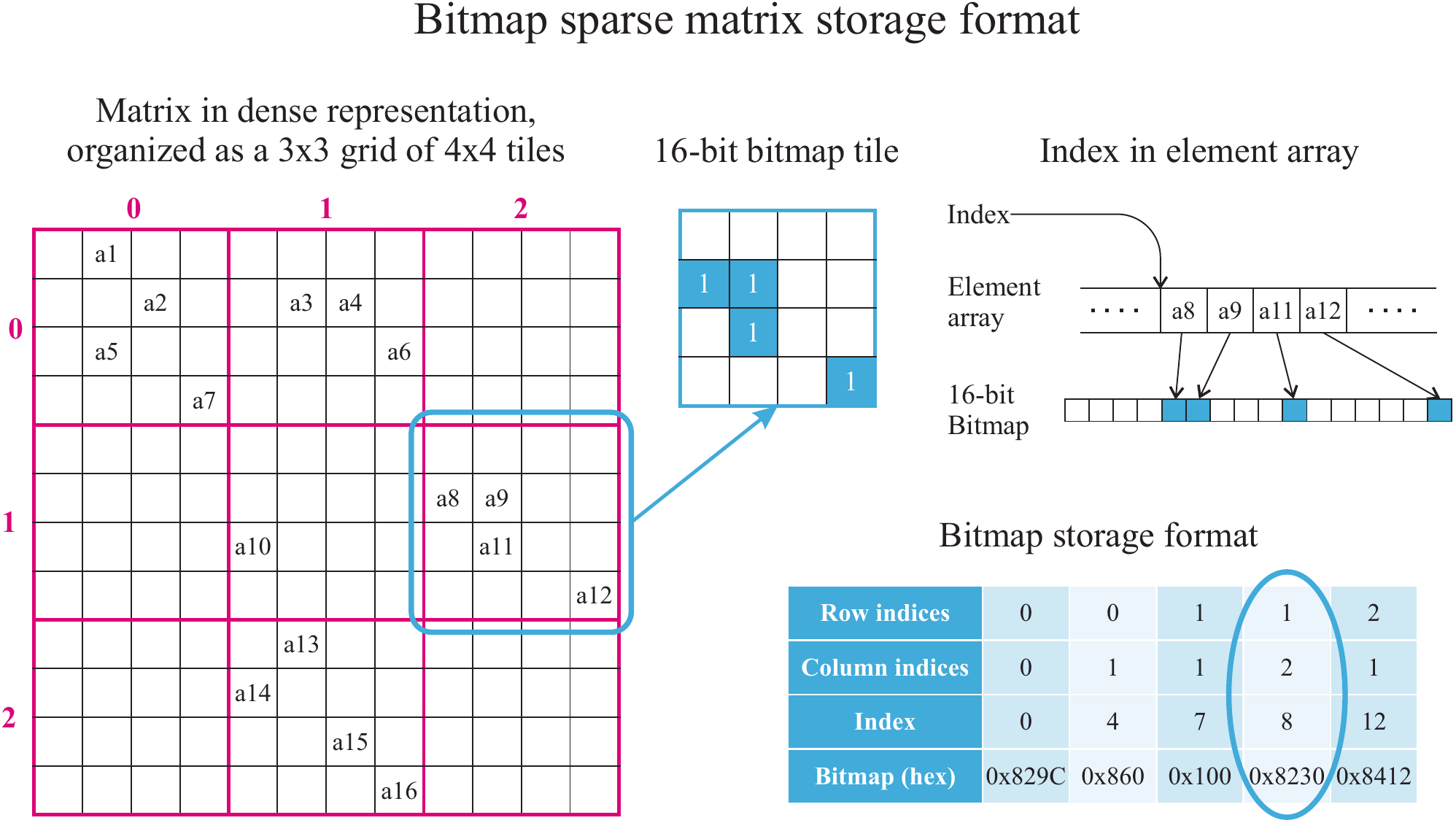}
	\caption{A $12 \PLH 12$ matrix in dense (left) and bitmap formats (bottom right). Tiles of $4 \PLH 4$ partition the $12 \PLH 12$ matrix in a $3 \PLH 3$ grid of tiles.  Non-zero elements $ a8,a9,a11,a12 $ of the circled tile are represented as \enquote{1} in the \emph{bitmap}. We store the NZ elements of the tile in consecutive locations in the element array. \emph{Index} points to the first element of the tile. On the bottom right of the figure, we circle the representation of the selected tile in bitmap format.}
	\label{fig:bitmap}                       
\end{figure*}

Fig.~\ref{fig:bitmap} shows how we convert a dense matrix to bitmap format. For simplicity we use $ 4 \PLH 4 $ tiles. We represent the positions of NZ elements as \enquote{1}s in the bitmap. We store four values for each tile that has one or more NZ elements: 1) row index like in COO format, 2) column index like in COO format,\replaced{}{and} 3) index into the element array, and 4) bitmap of the location of NZ elements in the tile.

\subsection{Sparse matrix-matrix multiplication} \label{sec:spgemm}

\begin{figure}[h!]
	\centering
	\includegraphics[width=0.65\linewidth]{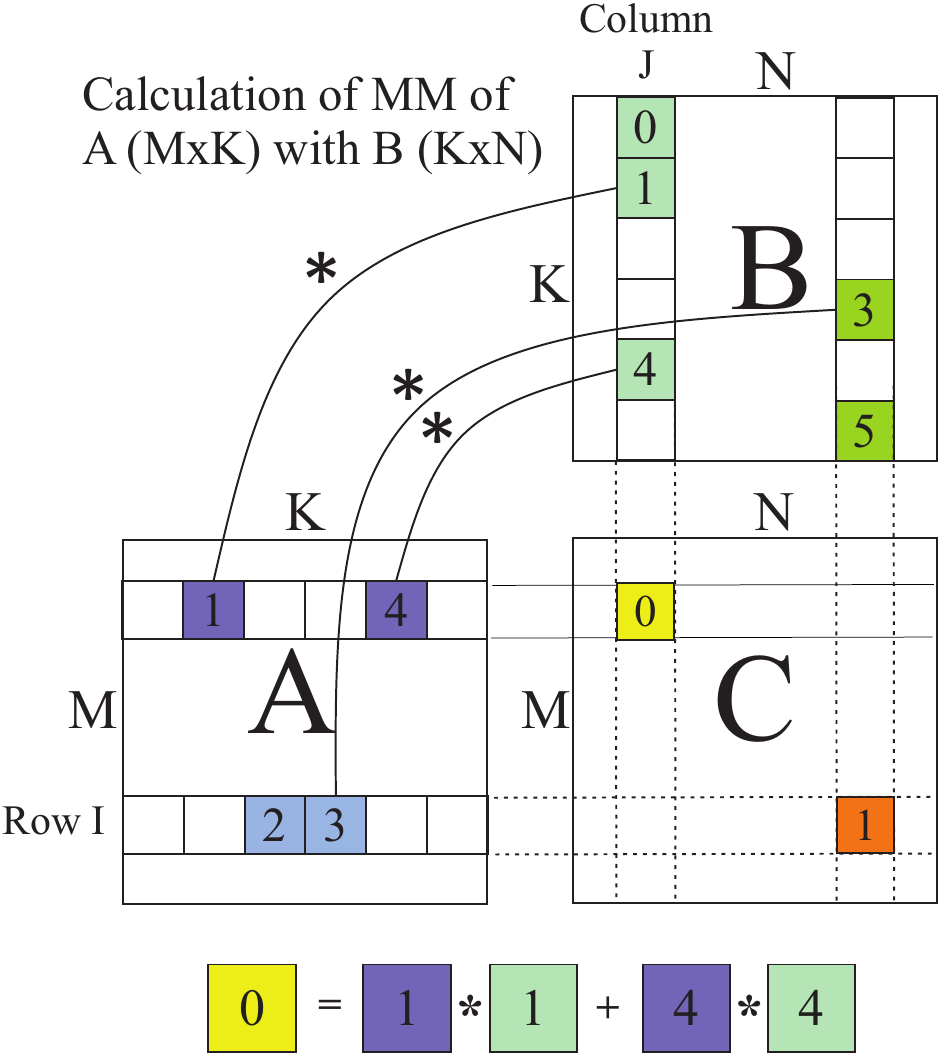}
	\caption{Example of the sparse matrix-matrix multiplication of two matrices \emph{A} and \emph{B} with dimensions $M \PLH K$ and $K \PLH N$ respectively. To find the output tile with coordinates $[I,~J]$, we multiply (MM) the corresponding NZ tiles of row \emph{I} of \emph{A} (dark blue) and column \emph{J} of \emph{B} (dark red).}
	\label{fig:mm}
\end{figure}

The general matrix multiplication (GEMM) has the form:
\begin{equation} \label{eq:gemm}
D = A \times B + C
\end{equation}
where \emph{A}, \emph{B}, \emph{C} are the input matrices and \emph{D} is the output.
In spGEMM, similarly to dense matrices, to get one element of the output, we need to multiply the NZ elements of one row of \emph{A} with the corresponding NZ elements of one column of \emph{B} and then accumulate the \emph{intermediate products} (i.e., calculate the inner product). The difference in spGEMM is that we multiply the corresponding elements only if the elements in the corresponding positions of the row of \emph{A} and the column of \emph{B} are both NZ and we accumulate only NZ products. The various ways to access the elements of \emph{A} and \emph{B} and perform MM are listed in \cite{matam_sparse_2012}.
The same multiplication method applies even if instead of elements we use tiles. In this case, the product of two corresponding tiles is their outer product (or equally MM).
We make two important observations.
First, \eqref{eq:gemm} takes the form
\begin{equation} \label{eq:gemm_acc}
C = A \times B + C
\end{equation}
when accumulating the tiles\replaced{, where C is both output and accumulator}{}. We use only the form of \eqref{eq:gemm_acc} for the rest of this work.
Second, the matrix Multiplication-Accumulation of small tiles is exactly what the TCUs were designed to do.
Fig.~\ref{fig:mm} shows an example of the $A*B$ sparse MM.

\subsection{Challenges of spGEMM} \label{sec:challenges}
Unlike other sparse matrix operations, in spGEMM both input and output are sparse. Therefore, it is very difficult to utilize the knowledge we infer from the sparsity structure of the input matrices to make arithmetic and memory optimizations. The reasons that make spGEMM more challenging than spMV (sparse matrix-vector multiplication) are three \cite{cusp_optimizing_2015, bhsparse_2015, winter_adaptive_2019, RMerge_2015, RMerge2_2018}:

First, data access is highly irregular because it depends on the sparsity structure of both matrices and the interaction of both structures. MM is not trivial for two reasons: 1) it requires access of possibly distant memory locations to load the inputs, and 2) it requires inserting the result to the output with possibly irregular access patterns.

Second, it is \replaced{computationally expensive}{very difficult or impossible} to know the size of the output before the actual MM.
According to Liu et al. \cite{bhsparse_2015}, there are four methods to estimate how much memory we need to allocate for the output. First, \replaced{the precise method, which makes }{precise. We make} an estimate that is very close \added{to} the actual size of the output, typically by \replaced{doing a }{}partial execution of the MM algorithm. Second,\replaced{ the }{}upper bound\replaced{ method, which typically uses}{. Usually,} as upper bound\replaced{}{, we use} the amount of intermediate products. Third, estimation using probability theory. 
Fourth, \replaced{progressive memory increase, which allocates}{by increasing the size of memory progressively, i.e., allocating} more memory if the previously allocated memory overflows. In all cases, at the end of spGEMM, we remove empty or unused entries from the allocated memory as necessary.

Third, load balancing. The sparsity structure of both matrices and the interaction of the structures determine the distribution of workload. NNZ elements in each row of the input or output may vary significantly, which makes it difficult to partition the workload among threads.

\subsection{Expand-Sort-Compress and CUSP} \label{sec:esc}
CUSP \cite{Cusp} is a library that specializes in sparse matrix operations. It is open-source and easily accessible on \replaced{GitHub}{github}. It is written in Thrust which makes it easy to read and port to other platforms. Therefore, it provides a good \enquote{boilerplate} to test our approach.

CUSP uses Expand-Sort-Compress ESC methodology. ESC performs the spGEMM in three steps \cite{Cusp, bell_exposing_2012}. First, Expand. \replaced{ESC multiplies}{We multiply} each NZ element $a_{i,j}$ of \emph{A} with all NZ elements of row $ B(j,:) $ of \emph{B} to get the intermediate products (no accumulation in this step) \cite{matam_sparse_2012}. Second, Sort. \replaced{ESC sorts}{We sort} the intermediate products of the previous step so that products that correspond to the same element of \emph{C} are in consecutive positions. Third, Compress. \replaced{ESC calculates}{We calculate} each element of \emph{C} by accumulating all respective products, which are in consecutive positions, thanks to the sorting step.

\subsection{Real number representation in digital computer systems} \label{sec:reals}
Computer systems have to store real numbers in bit representation. Floating point numbers are a common representation. The location of the decimal point and the number of bits determines the precision and range of the represented numbers. We denote the 32-bit representation as \emph{fp32}, whereas the 16-bit as \emph{fp16}. In contemporary systems, typically, we use floating point numbers as defined in IEEE 754 technical standard \cite{ieee2008754}. Usually, the fewer the number of bits, the faster the processing of the numbers is.

\subsection{Tensor Core Units} \label{sec:tcus}
\replaced{}{subsection CUDA}
NVIDIA, with the latest generations of Graphical Processing Units (GPUs) \replaced{}{(Turing architecture }\cite{turing_arch}\replaced{}{)}, brought Tensor Cores to the mainstream market. Nvidia TCUs\replaced{}{,} are ASICs that have the purpose of accelerating MM. Therefore, our work on spGEMM has significant benefits by properly adapting spGEMM to TCUs.
\replaced{}{We provide the necessary background of CUDA in this section.}

\replaced{}{subsubsection Cuda programming model}

\replaced{}{NVIDIA GPUs consist of several processors, which NVIDIA calls \replaced{Streaming}{Symmetric} Multiprocessors (SMs). Each SM includes a number of processing cores. GPUs of Turing architecture include cores for fp32 arithmetic, cores for integer arithmetic and tensor cores for MM \cite{turing_arch}.}

\replaced{}{In the finest granularity data are processed by \emph{threads}. Each SM can run many threads. An SM can run groups of 32 threads (called \emph{warps}) in \replaced{lockstep}{step-lock}. Threads are further organized \replaced{as}{to} groups, which are called \emph{thread blocks}. Thread blocks can contain up to 1024 threads. The GPU schedules thread blocks to SMs, where the thread blocks stay until completion \cite{CUDAProgrGuide}.}

\replaced{}{The GPU employs a multi-level memory hierarchy. Input and output data resides in the larger but slower off-chip memory, whereas frequently used data remain \replaced{on}{to} the faster but smaller on-chip memory. In CUDA, the off-chip memory is called \emph{global memory}. There are two types of on-chip memory: 1) \emph{L1 cache}, and 2) \emph{shared memory}. L1 is hardware managed, whereas shared memory is software-managed.}

\begin{figure}[h!]
	\centering
	\includegraphics[width=0.7\linewidth]{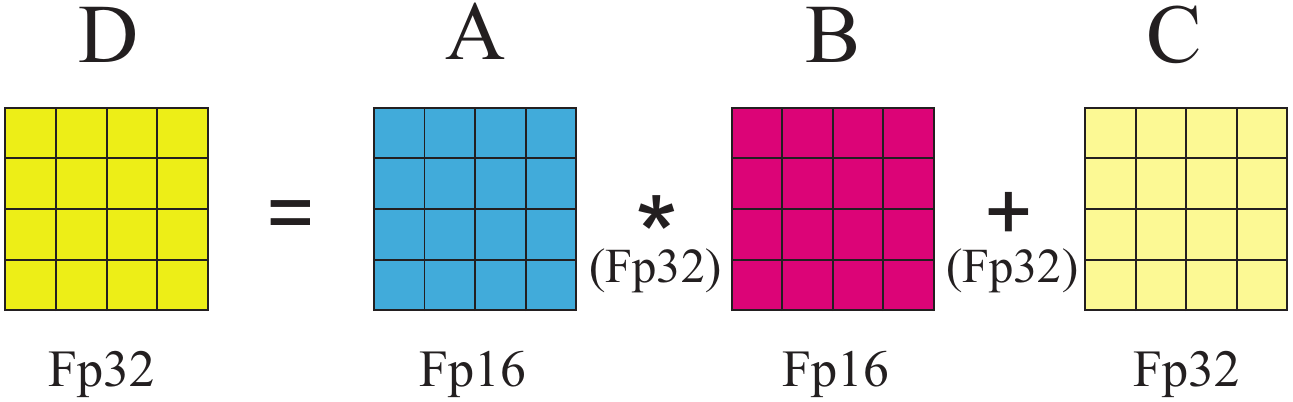}
	\caption{Mixed precision with CUDA TCUs. Inputs are stored in fp16, whereas the output and addend are stored in fp32. The multiplication and addition are performed in full precision.}
	\label{fig:mixed}
\end{figure}

TCUs\replaced{}{ specialize on MM and} mainly target deep learning, which is not very demanding precision-wise. Therefore, to accelerate MM, TCUs usually work with lower precision number representation (16-bit or less). However, fp16 or lower precision is detrimental to the output because precision and range of fp16 numbers can be insufficient when dealing with physical problems. To rectify this problem NVIDIA provides mixed precision functionality. Mixed precision allows to mix numbers of different precision. The two defining characteristics of the mixed precision implementation of NVIDIA are \replaced{as follows}{two}. First, although inputs \emph{A} and \emph{B} are in fp16 precision, their multiplication happens in full precision. Second, the product is stored as fp32 to accumulators \emph{C} and \emph{D} \cite{turing_arch}. Fig.~\ref{fig:mixed} shows the two characteristics. Markidis et al. and Haidar et al. \cite{markidis_nvidia_2018,haidar_harnessing_2018} evaluate the performance and precision of GEMM and linear equation solving using the mixed precision mode of TCUs. They show that TCUs can be used in other physical problems, outside deep learning. We use mixed precision functionality to extend the applicability of our work to non deep learning workloads. 

\section{Overview of our technique} \label{sec:overview}
Our TCU-based sparse GEMM technique alters ESC methodology in order to efficiently work with tiles. To that end, tSparse\replaced{}{,} creates a task list which delivers tiles to TCUs to accelerate the MM of the tiles.
\ore{At this point it is important to make a distinction between the terms \emph{elements} and \emph{tiles}. \emph{Elements} represent a single real or integer number, whereas \emph{tiles} represent a group of elements (Section \ref{ssec:bitmap}).}
This section gives a high level overview of the components of tSparse and how they relate \replaced{to}{with} each other.

\subsection{Creating the task list and allocating memory for tiles}
%

tSparse creates the task list in two steps.
First, tSparse determines which tiles of \emph{A} will be multiplied with which tiles of \emph{B} in order to get each tile of \emph{C}.
Second, unlike the expansion phase of ESC, tSparse does not immediately multiply the corresponding tiles of \emph{A} and \emph{B}. Instead, tSparse creates a task list that holds only the locations of the corresponding tuples of \emph{A} and \emph{B}. The entries of the task list are sorted so that pairs that correspond to the same tile of \emph{C} are in consecutive positions.
With the help of the task list, tSparse also estimates the number of \emph{C} tiles in the output and allocates memory accordingly.




\subsection{Counting kernel}
Dense matrices contain as many elements as the product of their dimensions. However, in sparse matrix-matrix multiplication, the NNZ of the output depend on the structure of the two input matrices and the interaction of these structures throughout the MM. Therefore, we know the exact size of the output only \emph{after} the matrices are multiplied. \ore{However, i}n order to allocate memory for storing the result of the MM we first need to know the exact size to allocate \ore{because} GPU kernels cannot \replaced{easily}{} reallocate memory during their execution. \ore{The purpose of the counting kernel is to make an estimation of the size of memory we need to allocate in order to store the element array of \emph{C}, }\emph{before} calling the multiplication kernel.

\ore{Non-blocking spGEMM approaches allocate memory only for elements. tSparse allocates memory for elements, in addition to memory for tiles.} However, allocating memory for elements is not as simple as calling a few parallel primitives because tile \ore{multiplication is a sparse MM}. The sparsity structure of both \ore{input} tiles and the interaction of the structures define the memory allocation requirements.

In our implementation, we use the precise method (Section \ref{sec:challenges}) to get an estimate of the \emph{count} of elements the output has\ore{, i.e., t}he counting kernel is a partial implementation of the multiplication kernel. The counting kernel requires shorter execution time than the multiplication kernel because it neither loads nor stores any elements. Instead, this kernel uses the bitmap to put zeros and ones in the place of the elements and \enquote{simulates} the MM. We allocate memory equal to the NNZ of the output of the simulated MM.

\subsection{Multiplication kernel}
Once we know how much memory to allocate for \ore{tiles} and elements we use the multiplication kernel to multiply and accumulate all pairs of tiles of \emph{A} and \emph{B} that correspond to each \emph{C} tile.

\subsection{Putting everything together}
In order to perform the matrix multiplication of \emph{A} with \emph{B}, we 1) determine which products need to be accumulated for each tile of \emph{C}, and 2) allocate memory for the \ore{tiles} of \emph{C} and the element array. Using the task list and the counting kernel we determine the memory allocation size for the \ore{tiles} of \emph{C} and the element array, respectively. Subsequently, the multiplication kernel has everything it needs to multiply \emph{A} with \emph{B}.

We expect three benefits. First, by placing both the multiplication and accumulation steps of MM in the multiplication kernel we can use TCUs for MM. By moving MM to TCUs, the computational heavy MM is no longer a bottleneck of the spGEMM algorithm. Second, the use of bitmap format reduces memory consumption because one row index and one column index represents up to 64 elements \cite{bmsparse_2018}. Third, by grouping elements to tiles, we reduce the amount of values we have to manipulate and therefore there are additional performance benefits (e.g., \ore{fewer} values to sort during sorting phase of ESC).
	
\section{Implementation details} \label{sec:impl}
This section describes the three main parts of our approach, tSparse, in detail: 1) the creation of the task list, 2) the estimation of how much memory to allocate for the\ore{ elements of the} output, and 3) the MM of the tiles. It also describes some smaller components of our implementation.

\subsection{Creating the task list and allocating memory for tiles}

The four main parts of our algorithm are: 1) finding corresponding tiles of \emph{A} and \emph{B}, 2) filtering of zero products, 3) creating a task list with entries that point to the tuples of \emph{A} and \emph{B}, and 4) estimating how much memory to allocate for the tuples of the output. In detail:

First, similarly to CUSP, tSparse uses parallel primitives from Thrust library to find the correspondence among tiles of \emph{A} and tiles of \emph{B}.
The \ore{main} difference with CUSP is that \ore{instead of using single elements we use tiles. This is the part of our methodology that is the same between CUSP and tSparse. We do not change it because it is efficient and accounts for less than 15\% of the total execution time of our spGEMM.}

Second, the algorithm that finds the correspondence among tiles of \emph{A} and \emph{B} considers the whole tile as a single value, i.e., the result is the same regardless if the $8 \PLH 8$ tile contains one or 64 elements. Therefore, the correspondence algorithm finds a correspondence \ore{between} tiles \ore{of \emph{A} and \emph{B}} even if the elements inside the two tiles are not corresponding. The resulting tile of the MM of such tiles is a tile with only zero elements. tSparse has a routine that removes this type of corresponding tiles by applying boolean arithmetic on bitmaps. This routine is fast as it does not compute MMs. The culling significantly lightens the workload of the rest of our methodology (fewer tiles to sort, multiply etc.).

Third, tSparse creates the task list. An important consideration is that in ESC methodology the MM and accumulation are in different steps, i.e., MM is in the Expand step and accumulation in the Compress step. However, to fully take advantage of the combined MAC operation of TCUs we need to perform both MM and accumulation of tiles in the same step. In order to put MM and accumulation in the same step we sort the \emph{locations} of the multiplicands, instead of sorting the intermediate products of the MM. By sorting the locations, we defer the MM step until after the Sorting step. Effectively, the locations of the multiplicands form a task list, of which each entry points to one tile of \emph{A} and one tile of \emph{B}.
Merging the expansion and compression steps and moving them after the sorting step significantly reduces both memory allocations and costly data movement to/from global memory. The reason is that tSparse does not store the intermediate products. Therefore, we save memory by not allocating memory for the 64-bit bitmaps, an undefined number of up to 64 elements and the corresponding indices in the element array. We reduce data movements to/from global memory by not storing the intermediate products before the sorting step and loading them for a second time after the sorting step.

Fourth, our algorithm counts how many of the intermediate products correspond to the same tile of \emph{C} using a segmented reduce parallel primitive on the sorted indices. We create an offset array from the prefix sum of the counted intermediate products. Our GPU kernels use the offset array for indexing purposes. The prefix sum also gives the total count of tuples in the output. We use this total count to allocate memory.\ore{

\paragraph{Sorting}
Sorting a long task list is computationally demanding, therefore we need advanced optimization strategies. tSparse works with sorted matrices. We use this knowledge to sort the workload of each row of \emph{A} separately. For this task, we use the segmented sort by Kaixi et al. \cite{hou2017fast}, which employs a hybrid sorting scheme based on row length.}

\subsection{Counting kernel}

The counting kernel works in four steps. First, it reads the bitmaps of \emph{A} and \emph{B}. Second, it creates tiles, wherein each element is set to \enquote{1} or \enquote{0} based on the corresponding position in the bitmap. Third, it multiplies and accumulates the tiles of \emph{A} and \emph{B} that correspond to each individual tile of \emph{C} using TCUs. Fourth, we count how many elements of the resulting tile are NZ with the \texttt{ballot} instruction. We repeat for all tiles of \emph{C} and accumulate the counts.
\ore{The counting kernel returns an array of which each value holds an estimation of how many elements each tile of \emph{C} has.}

Each tile contains only \enquote{1}s or \enquote{0}s, therefore: 1) the estimation of memory requirements \ore{can be more than} what is actually required because the counting kernel cannot account for \ore{possible numerical }cancellation (after the actual multiplication there is a compaction stage that removes empty entries from the allocated \emph{element array}), and 2) half precision is enough for executing the MM of the counting kernel because we work only with zero and NZ elements.

\subsection{Multiplication kernel} \label{ssec:mul_impl}
The multiplication kernel performs the actual multiplication and constructs the COO matrix of the output, i.e., it sets the row and column indices, the index and bitmap tuple and the elements of the element array. The multiplication kernel loads the actual elements from \emph{A} and \emph{B} and stores the result in the memory allocated by the counting kernel. We emphasize that we use only TCUs for the MM and not TCUs in addition to the standard CUDA cores.

There are two important considerations when multiplying the elements, which are real numbers.
First, fp16 arithmetic has a very limited representation range of numbers
, which we can easily exceed with multiplication. Therefore, we prefer the mixed precision functionality of TCUs.
Second, unlike the counting kernel where we have only positive numbers, when accumulating real numbers, elements get canceled as a result of \ore{numerical cancellation}. Many tiles may end up empty, something that the counting kernel, which acts on boolean elements, does not predict. For this reason, our multiplication kernel has the additional task of marking for removal \emph{tiles} that are empty.

\begin{figure}[h!]
	\centering
	\includegraphics[width=0.99\linewidth]{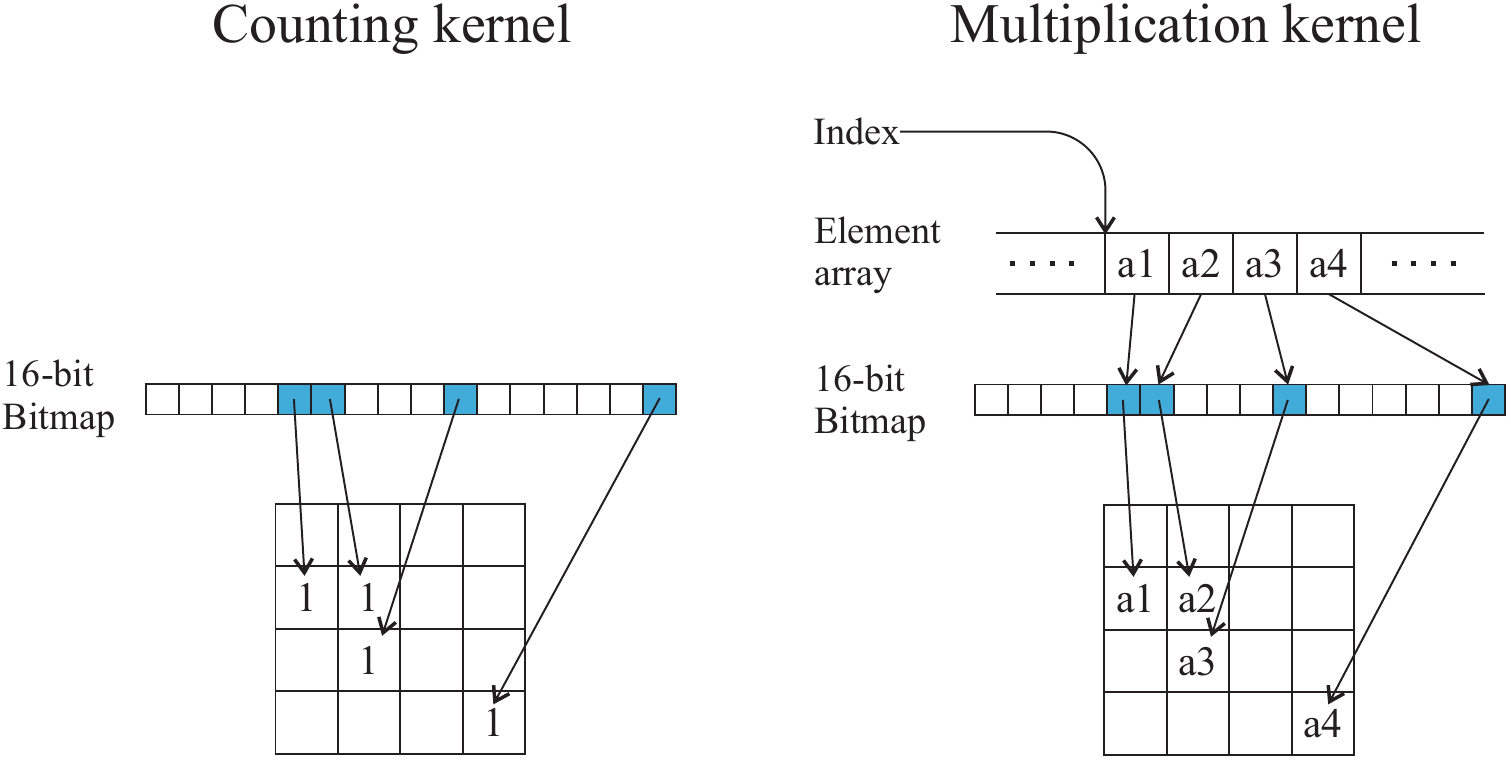}
	\caption{Comparison of counting and multiplication kernels. The counting kernel (left) places \enquote{1}s at the locations indicated by the bitmap. The multiplication kernel (right) loads the actual elements from memory and places them at the locations indicated by the bitmap.}
	\label{fig:kernels}
\end{figure}

Fig.~\ref{fig:kernels} compares element loading between the counting and multiplication kernels. The counting kernel does not need to load any actual element. It just creates \enquote{1}s based on the bitmap. The multiplication kernel, on the other hand, loads the elements and it places them according to the bitmap.

\subsection{Other components} \label{sec:impl_common}
\begin{figure}[h!]
	\centering
	\includegraphics[width=0.45\linewidth]{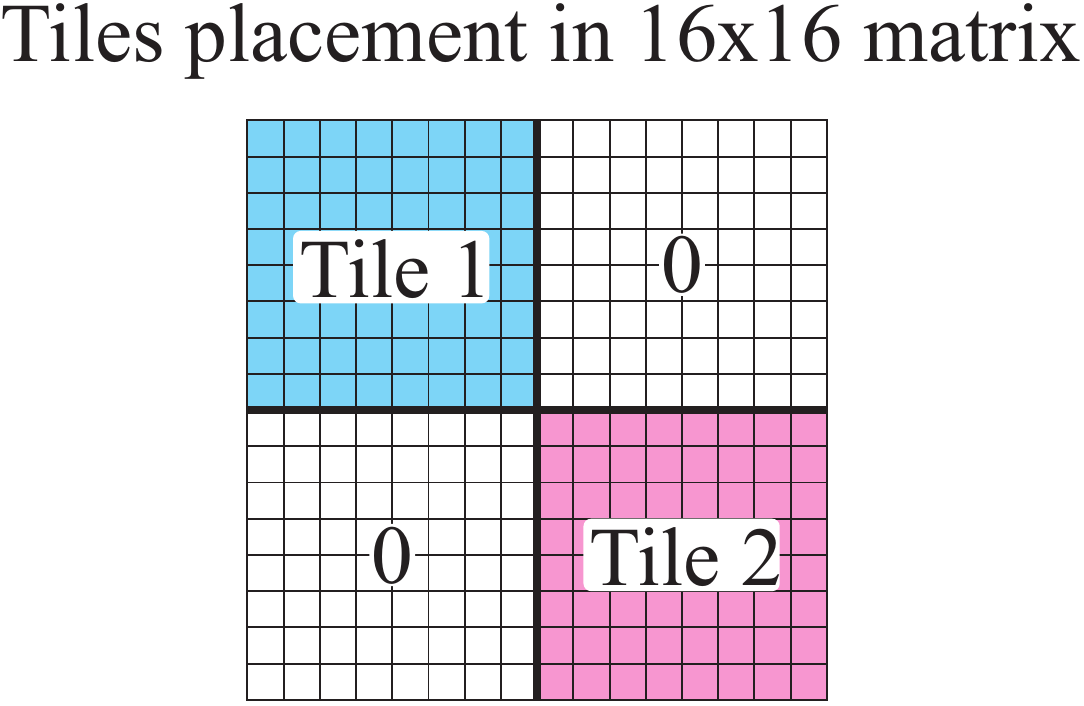}
	\caption{Placement of two $8 \PLH 8$ tiles in the $16 \PLH 16$ matrix configuration of TCUs.}
	\label{fig:tilesTCU}
\end{figure}

\paragraph{Arrangement of tiles in the TCUs}
NVIDIA does not provide an Application Programming Interface (API) for using TCUs with small $8 \PLH 8$ tiles \cite{CUDAProgrGuide}.
TCUs execute MM on 256 elements at a time. However, our tiles have a size of $8 \PLH 8$, which means that a large part of the TCU remains unused. Although a TCU does not have to be fully loaded in order to get performance benefits, we can fit two tiles in a single TCU. $16 \PLH 16$ is the only supported matrix configuration that can fit two $8 \PLH 8$ tiles \cite{CUDAProgrGuide}. To put two tiles in the same TCU two steps are necessary. First, we initialize the $16 \PLH 16$ matrix to zero. Second, the tiles must be placed in a diagonal of the $16 \PLH 16$ matrix (Fig.~\ref{fig:tilesTCU}). If the tiles were not in the diagonal, but instead side-by-side (top-bottom), the same row (column) of the $16 \PLH 16$ matrix would have elements of two unrelated tiles, which would mix the inner products (of rows of \emph{A} with columns of \emph{B}) of the first tile with the inner products of the second tile. \ore{The API forces loading to/from TCUs through shared memory. We find the internal layout of the registers of the TCUs (fragments) and we access them directly instead. Loading through registers, which are faster than shared memory, we minimize data movement and increase the performance of MM.}

\paragraph{Load balancing}
Both counting and multiplication kernels calculate an inner product of tiles (Section \ref{sec:spgemm}). The number of intermediate products required for each tile of \emph{C} is different, because the number depends on the sparsity structure of \emph{A} and \emph{B}. Therefore the workload for the computation of each tile of \emph{C} is different, leading to thread blocks with different amounts of work. Thread blocks with different execution times create imbalance among the SMs of the GPU. To tackle this issue, we assign each \ore{pair of tiles} of \emph{C} to a different thread block. When a thread block finishes and releases the resources, the scheduler of the SM schedules another thread block to take its place. Therefore, SMs take blocks according to their needs and stay fully occupied until completion of the MM. 

%
%
%

\begin{figure*}[h!]
	\centering
	\includegraphics[width=0.75\linewidth]{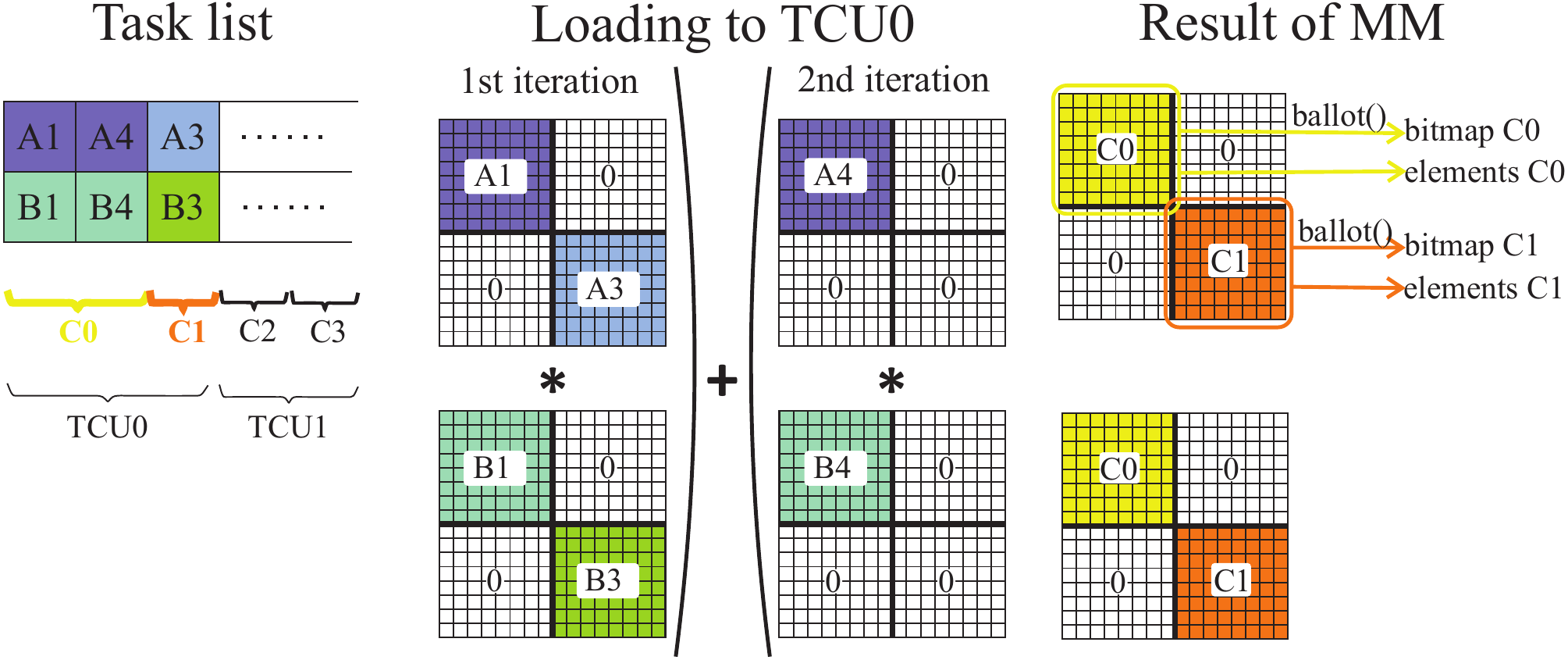}
	\caption{Detail of the multiplication kernel. Each TCU calculates two tiles of the output \emph{C}. If the two output tiles require a different number of addends, we use \enquote{0}s to make them match. Finally, the output of the TCU gives us the respective output elements of the two tiles and their bitmaps (using ballot operations).}
	\label{fig:multiplication_kernel}
\end{figure*}

\paragraph{Compaction of zeros}
MM creates zero elements because of \ore{numerical }cancellation. To store the output in a strictly sparse format we need to remove all zeros. Therefore, we need a way to detect zeros and remove them or, in other words, \emph{compact} the arrays that hold the elements and the tuples. For the element array, which is just an array, we use a compaction parallel primitive from Thrust. For the array that holds the tuples, we first mark empty tiles in the multiplication kernel (see Section \ref{ssec:mul_impl}).\ore{
	
\subsection{Putting everything together}
	
Algorithm \ref{alg:tsparse} summarizes tSparse. First, tSparse creates a task list (lines \ref{alg:tsparse:task_list:start}-\ref{alg:tsparse:task_list:end}). Second, it estimates how much memory to allocate for tiles (lines \ref{alg:tsparse:task_list:alloc_tiles:start}-\ref{alg:tsparse:task_list:alloc_tiles:end}). Third, it estimates how much memory to allocate for elements (lines \ref{alg:tsparse:counting:start}-\ref{alg:tsparse:counting:end}). Fourth, it multiplies the matrices (line \ref{alg:tsparse:multiplication}). Finally, it compacts zero elements (line \ref{alg:tsparse:compaction:elements}) and empty tiles (line \ref{alg:tsparse:compaction:tiles}).

\begin{algorithm}[h!]
	\caption{Pseudocode for tSparse}
	\label{alg:tsparse}
	\begin{algorithmic}[1]
		\FORALL{NNZ tiles $ A[i,j] $ \textbf{in} $  A[:,:] $} \label{alg:tsparse:task_list:start}
		\FORALL{NNZ tiles $ B[j,k] $ \textbf{in} $ B[j,:] $}
		\STATE $task\_list \gets \{row\_ptr(A[i,j]), col\_ptr(B[j,k])\}$
		\ENDFOR
		\ENDFOR
		\STATE \textsc{FilterTilesWithZeroProduct}($ task\_list $)
		\STATE \textsc{SortByKey}($B_{cols}[task\_list], task\_list$)
		\STATE \textsc{SortByKey}($A_{rows}[task\_list], task\_list$) \label{alg:tsparse:task_list:end}
		\STATE $tile\_count \gets 0$ \label{alg:tsparse:task_list:alloc_tiles:start}
		\FORALL{$ c $ \textbf{in} $ task\_list $}
		\IF{$ C[A_{rows}[c],B_{cols}[c]] $ is unique}
		\STATE $tile\_count \gets tile\_count + 1$
		\ENDIF
		\ENDFOR
		\STATE \textsc{AllocateMemGPU}($ tile\_count $) \label{alg:tsparse:task_list:alloc_tiles:end}
		\STATE $element\_count \gets$ \textsc{CountingKernel}($ task\_list $) \label{alg:tsparse:counting:start}
		\STATE \textsc{AllocateMemGPU}($ element\_count $) \label{alg:tsparse:counting:end}
		\STATE $C_{tiles}, C_{elements} \gets$ \textsc{MultiplicationKernel}($ task\_list $) \label{alg:tsparse:multiplication}
		\STATE \textsc{CompactZeroElements}($ C_{elements} $) \label{alg:tsparse:compaction:elements}
		\STATE \textsc{CompactEmptyTiles}($C_{tiles}$) \label{alg:tsparse:compaction:tiles}
	\end{algorithmic}
\end{algorithm}
	
Figure \ref{fig:multiplication_kernel} illustrates details of the multiplication kernel (Algorithm \ref{alg:tsparse}, line \ref{alg:tsparse:multiplication}). In order to get a tile \emph{C} of the output, we have to accumulate a varied number of products (MMs of tiles of \emph{A} and \emph{B}), e.g., \replaced{$ C0 = A1 \times B1 + A4 \times B4  $ and  $ C1 = A3 \times B3 $ for the example of Fig. \ref{fig:mm} (note the color code of tiles)}{$ C0 = A0_0 \times B0_0 + A0_1 \times B0_1  $ and  $ C1 = A1_0 \times B1_0 $}. Each TCU processes two tiles, e.g., TCU0 calculates \emph{C0} and \emph{C1}. However, \emph{C1} has fewer addends. Therefore, when we load the second addend of \emph{C0}, we use \emph{0}s in place of the second addend of \emph{C1}. From the final result, we extract the bitmaps (using \texttt{ballot}) and elements of the output.}

\section{Evaluation Methodology} \label{sec:eval}
We test our approach, tSparse, on \ore{two systems: 1) Intel i9-9900@3.6GHz CPU and NVIDIA Titan RTX GPU, and 2) Intel i7-8700@3.2GHz CPU and NVIDIA RTX 2070 GPU. Both GPUs are of the Turing architecture \cite{turing_arch}}. We use CUDA SDK v10.\ore{2} and the accompanying parallel primitives library, Thrust \cite{CUDAProgrGuide}, for our GPU code.

We compare tSparse with cuSPARSE from CUDA Toolkit \cite{CUDAProgrGuide}, CUSP \cite{Cusp}, RMerge2 \cite{RMerge2_2018}, Nsparse \cite{nagasaka_high-performance_2017}, AC-SpGEMM \cite{winter_adaptive_2019}, spECK \cite{parger_speck_2020}. In addition, to confirm the benefit of using TCUs, we create one implementation of tSparse without TCUs and we name it \emph{nonTCU}. In nonTCU, we use the same method as \cite{bmsparse_2018} to multiply the tiles (Algorithm \ref{alg:tile_multiply}). \replaced{We note that we include the time of memory allocations in the execution time of AC-SpGEMM.}{}

\begin{algorithm}
	\caption{Matrix multiplication of two $8 \PLH 8$ tiles without TCUs}
	\label{alg:tile_multiply}
	\begin{algorithmic}[1]
		\STATE $tid$ \COMMENT{The id of a thread}
		\STATE $i \gets 0$
		\WHILE{$i < 8$}
		\STATE $C\_tile[tid] = C\_tile[tid] + A\_tile[(tid/8)*dim+i] * B\_tile[i*8+mod(tid,8)]$
		\ENDWHILE
	\end{algorithmic}
\end{algorithm}

To evaluate the performance of tSparse, we perform the $A*A$ MM, which has the benefit that both matrices have the same sparsity structure. The same structure makes it easier to make observations.
\replaced{In order to facilitate the presentation of the performance of tSparse, we select a subset of 16}{We select} matrices from SuiteSparse Matrix Collection
\cite{suitesparse} for our dataset. All selected matrices are square, as our $A*A$ problem dictates. We select matrices which have elements in the fp16 range.

\begin{table*}[h!]
	\centering
	\resizebox{0.85\textwidth}{!}{
		\begin{tabular}{@{}lrrrrrrr@{}}
			\toprule
			Matrix name  &    \makecell[r]{Dimensions \\ (square)} &  NNZ(A) &   NNZ(C) & NNZ(\textoverline{C}) & NNZ(C\textsubscript{tiles}) & NNZ(\textoverline{C}\textsubscript{tiles}) &  \makecell[r]{bitmap density \\ (median, mean, std)} \\ \midrule
			mc2depi        & 525825  & 2100225 & 5245952  & 8391680   & 718228  & 1364149  & 7, 6.4, 4.9    \\
			webbase-1M     & 1000005 & 3105536 & 51111996 & 69524195  & 2546355 & 4327469  & 6, 5.6, 5.3    \\
			cage12         & 130228  & 2032536 & 15231874 & 34610826  & 2945653 & 9295217 & 3, 4.5, 4.1    \\
			dawson5        & 51537   & 1010777 & 3616737  & 21284355  & 219077  & 1515543  & 6, 9.5, 9.1 \\
			lock1074       & 1074    & 51588   & 134676   & 2752056   & 3050    & 19520   & 32, 31.4, 17.3 \\
			patents\_main  & 240547  & 560943  & 2281308  & 2604790   & 2089143 & 2511808 & 1, 1.0, 0.2    \\
			struct3       & 53570   & 1173694 & 3400384  & 26704476  & 146007  & 543125  & 21, 18.8, 9.8 \\
			wiki-Vote      & 8297    & 103689  & 1831112  & 4542805   & 526421  & 3058660  & 1, 1.4, 1.0    \\ \midrule
			bcsstk30       & 28924   & 2043492 & 8946070  & 173481412 & 252076  & 1627326  & 23, 25.9, 17.9 \\
			nemeth21       & 9506    & 1173746 & 2578720  & 146859992 & 47341   & 523549  & 59, 47.0, 21.1 \\
			pcrystk03      & 24696   & 1751178 & 7240266  & 129128312 & 212471  & 1514965  & 15, 23.4, 18.8 \\
			pct20stif      & 52329   & 2698463 & 10016951 & 154237335 & 323396  & 1770466  & 17, 23.4, 18.1   \\
			pkustk06       & 43164   & 2571768 & 10596384 & 179924544 & 451380  & 2336176  & 16, 19.6, 12.9 \\
			pli            & 22695   & 1350309 & 8548665  & 99698581  & 292851  & 2262407  & 14, 15.9, 11.5 \\
			net50          & 16320   & 945200 & 40622452  & 79727280  & 1037234 & 8150260   & 8, 8.6, 11.1 \\
			web-NotreDame  & 325729  & 1497134 & 16801350 & 64593748  & 693759  & 1589613  & 3 , 6.5, 10.7 \\ \bottomrule
		\end{tabular}
	}
	\caption{Matrix characteristics. We list the size of the matrix (number of rows/columns), the number of non-zeros of: the input (NNZ(A)), the output (NNZ(C)), the intermediate matrix (NNZ(\textoverline{C}), the number of tiles (NNZ(C\textsubscript{tiles})), and the number of tiles of the intermediate matrix (NNZ(\textoverline{C}\textsubscript{tiles})), and the density of the tiles of the input matrix. The upper part corresponds to matrices that are commonly used in the literature, the bottom part to matrices we selected based on our criteria.} \label{tab:data}
\end{table*}

Table \ref{tab:data} shows the characteristics of our dataset. We denote a matrix stored in bitmap storage format as \emph{C\textsubscript{tiles}}. \replaced{For non-tiling (i.e., single element) approaches, we denote as \emph{\textoverline{C}} the total amount of intermediate products. Similarly, for our tiling approach, we denote as \emph{\textoverline{C}\textsubscript{tiles}} the total amount of intermediate products, where each product is a MM between tiles.}{We denote the matrices of the intermediate products as \emph{\textoverline{C}\textsubscript{tiles}} and \emph{\textoverline{C}}, for bitmap and non-bitmap storage formats respectively (although tSparse does not actually store the intermediate products).} $NNZ(\cdot)$ of a matrix denotes the NNZ values of the matrix. \replaced{Finally, from the second and third columns of Table \ref{tab:data} derives the average row size of the input, which is defined as $NNZ(A) / Dimensions$, and we denote as $\widehat{RowA}$.}{}

We divide our dataset into two parts. The first part (upper half of Table \ref{tab:data}) consists of matrices that other works use \cite{cusp_optimizing_2015,bhsparse_2015,winter_adaptive_2019,RMerge_2015,RMerge2_2018,bmsparse_2018,nagasaka_high-performance_2017,cusparse_presentation_2012,anh_balanced_2016,kunchum_improving_2017}. The second part (lower half of Table \ref{tab:data}) consists of matrices that we select after taking into consideration the two criteria that derive from the \replaced{analysis}{performance results} in Section \ref{sec:criteria}: \replaced{1) $NNZ(A) > 300000$, and 2) $\widehat{RowA} > 42$}{1) $NNZ(A) > 1000000$, and 2) \ore{$NNZ(\overline{C}) / NNZ(\overline{C}\textsubscript{tiles}) > 9$}}.

\replaced{}{We collect two types of measurements: 1) the execution time of the spGEMM, and 2) the speedup of tSparse over the other implementations.}.


\section{Results and Analysis} \label{sec:results}






\replaced{}{In this work, we use a tiling approach to group NZ elements. Tiles are suitable for MM with TCUs. To show the performance benefits we compare tSparse with other approaches from the state-of-the-art, as well as, with one version of tSparse that does not use TCUs.}\replaced{}{To study the effect of our optimizations, we execute the $A*A$ spGEMM and \replaced{}{we} measure the execution time and the speedup of tSparse over the other approaches.}

\replaced{In this section, we find the criteria that define when tSparse is the recommended spGEMM approach using all qualified matrices from SuiteSparse. Then, }{}we collect four types of measurements using the matrices in Table \ref{tab:data}: 1) \replaced{the speedup of tSparse}{the speedup of tSparse over the other implementations}, 2) the execution time \replaced{breakdown}{of the spGEMM}, \replaced{3) the numerical precision, and 4) the memory consumption}{}.

\subsection{Finding the selection criteria} \label{sec:criteria}
\replaced{We measure the execution time of cuSPARSE, CUSP, RMerge2, Nsparse, AC-SpGEMM, spECK, nonTCU and tSparse on a collection of matrices which we create from SuiteSparse. For this collection, we keep matrices that meet the following four conditions: 1) the input matrix has more than 10000 NNZ elements, 2) the input is square with real numbers in the fp16 range, 3) the bitmap density of the input is more than one, and 4) matrices for which all approaches that participate in the comparison return correct results. We then define criteria based on the characteristics of the matrices. These criteria facilitate the selection of the most appropriate spGEMM approach among tSparse and the other approaches.}{}

\replaced{We find the criteria in two steps. First, we analyze how each approach works in order to identify which matrix characteristics affect the performance of said approach. Second, we adjust the criteria to our matrix collection manually. Alternatively, we could use a machine-learning method to find the criteria \cite{xie2019ia}, which we leave for future work. We make six major observations.}{}

\replaced{First, tSparse generally outperforms cuSPARSE in the larger matrices of our collection ($NNZ(A) > 200000$). This happens probably because cuSPARSE does not have sufficient shared memory to store the hash tables and therefore global memory traffic increases. Nevertheless, the size of the matrix is not the only thing that decides the performance of cuSPARSE. To the best of our knowledge, the performance of cuSPARSE also depends on the sparsity structure of the matrix because the structure affects the number of hash conflicts.}{}

\replaced{Second, tSparse performs better than CUSP when the number of intermediate products of tiles is smaller than the number of intermediate products of elements of CUSP, i.e., $NNZ(\overline{C}\textsubscript{tiles}) < NNZ(\overline{C})$.
Otherwise, CUSP, which handles only single elements, is faster than tSparse which has an additional overhead for handling tiles.}{}
	
\replaced{Third, tSparse is faster than Nsparse when the average row size of \emph{A} ($\widehat{RowA}$) is greater than 42 and $NNZ(A) > 100000$. Nsparse follows a hash table approach (Section \ref{ssec:hash}). Therefore, larger rows possibly create more hash conflicts and hash tables are difficult to keep in shared memory (instead of the slower global memory) with large rows. We also note that web graph matrices (webbase-1M and web-NotreDame) have high irregularity, i.e., although the average row size is small (3.1 and 4.6 respectively), there are rows with thousands of elements (maximum 4700 and 3445 respectively). Therefore they are also taxing for Nsparse.}{}
	
\replaced{Fourth, RMerge2 shows behavior similar to Nsparse. tSparse has better performance than RMerge2 when $\widehat{RowA} > 42$. To the best of our knowledge, RMerge2, which follows a hybrid approach (Section \ref{ssec:hybrid}), uses faster kernels when the row size is smaller than the size of a CUDA warp \cite{RMerge2_2018}. We note that the low recall score is owed to an instability of RMerge2 (possibly caused by a driver issue) that, in a seemingly random manner, decreases the performance.}{}

\replaced{Fifth, AC-SpGEMM is an ESC-based approach, like CUSP. Therefore, a similar criterion applies to AC-SpGEMM. The only difference is that, as AC-SpGEMM has much better performance than CUSP, the criterion is much stricter, i.e., $9 * NNZ(\overline{C}\textsubscript{tiles}) < NNZ(\overline{C})$.}{}

\replaced{Sixth, spECK is a hash table approach and has similar behavior to Nsparse. spECK generally performs better than Nsparse, therefore we select a criterion similar to Nsparse's but stricter, i.e., $\widehat{RowA} > 42$ and $NNZ(A) > 300000$.}{}

\begin{table*}[]
	\centering
	\resizebox{0.65\textwidth}{!}{
		\begin{tabular}{@{}lcrrr@{}}
			\toprule
			Approach  &                                  Condition                                   & \makecell[r]{Predictions \\ /Collection size}  & Precision & Recall \\ \midrule
			cuSPARSE  &                          NNZ(A) \textgreater 200000                          &           122/260 &      0.75 & 0.84   \\
			CUSP      & NNZ(\textoverline{C}) / NNZ(\textoverline{C}\textsubscript{tiles}) $ \geq 1$ &           260/260 &      0.98 & 1      \\
			RMerge2   &                   $\widehat{RowA} > 42$ AND NNZ(A) $> 100000$                    &            53/233 &      0.92 & 0.33   \\
			Nsparse   &                   $\widehat{RowA} > 42$ AND NNZ(A) $> 100000$                    &            53/233 &      0.75 & 0.74   \\
			AC-SpGEMM &   NNZ(\textoverline{C}) / NNZ(\textoverline{C}\textsubscript{tiles}) $> 9$   &           120/233 &      0.89 & 0.86   \\
			spECK     &                   $\widehat{RowA} > 42$ AND NNZ(A) $> 300000$                    &            45/234 &      0.78 & 0.8    \\ \midrule
			All       &                   $\widehat{RowA} > 42$ AND NNZ(A) $> 300000$                    &            45/233 &      0.76 & 0.85   \\ \bottomrule
		\end{tabular}
	}
	\caption{The criteria that define when tSparse is faster in comparison to other approaches. This table evaluates the performance of each condition using the precision and recall metrics from the classification theory. We note that the collection size varies to account for the incorrect results of each approach.} \label{tab:criteria}
\end{table*}

\paragraph{Global criterion}
\replaced{In summary, a range of matrix characteristics define the performance of each spGEMM approach. To help with selecting tSparse over other approaches, we define two criteria that generally work well to show when tSparse is the most appropriate approach: 1) $NNZ(A) > 300000$, and 2) $\widehat{RowA} > 42$. Table \ref{tab:criteria} summarizes the various criteria. We note that tSparse performs better on RTX 2070 than on Titan RTX, for the reasons we describe in Section \ref{sec:speedup}. Therefore, we can relax the criterion for non-high-end GPUs as follows: $\widehat{RowA} > 21$. Finally, we note that the first criterion may limit the suitability of our approach for applications that reduce the size of input matrices, e.g., AMG.}{}

\subsubsection{Evaluating the performance of the criteria}
\replaced{We evaluate the speedup of tSparse over the other spGEMM approaches after applying our global criterion to the matrix collection. Table \ref{tab:speedup} shows the speedup of tSparse over the other spGEMM approaches when applying the global criterion.}{}

\replaced{We make three major observations.
	First, tSparse is faster than all other approaches by $1.46 \PLH - 36.93 \PLH$, which confirms that our global criterion is reliable.
	Second, the speedup of all approaches follows the same trends for both GPUs. The RTX 2070 GPU performs better w.r.t. execution time, with more relaxed criterion, as the CPU does not bottleneck the execution time as much as on Titan RTX (Sections \ref{sec:breakdown}, \ref{sec:speedup}).
	Third, the \emph{Minimum} columns of Table \ref{tab:speedup} show speedups $<1$. This happens because of the wrong predictions of our proposed criterion (false positives).}{}

\begin{table*}[]
    \centering
    \resizebox{0.65\textwidth}{!}{
        \begin{tabular}{@{}lrrrrrr@{}}
            \toprule
             & \multicolumn{6}{c}{Speedup with tSparse} \\ \cmidrule(lr){2-7}
             \multirow{2}{*}{Approach}  & \multicolumn{3}{c}{Titan RTX ($\widehat{RowA}<$ 42)} & \multicolumn{3}{c}{RTX 2070 ($\widehat{RowA} <$ 21)} \\ \cmidrule(lr){2-4} \cmidrule(lr){5-7}
             & Gmean & Minimum & Maximum & Gmean & Minimum & Maximum\\ \midrule
            cuSPARSE  & 2.8              & 0.65        & 11.58    & 2.98        & 0.78        & 9.37     \\
            CUSP      & 36.93            & 4.17        & 132.74   & 26.25       & 3.55        & 113.5    \\
            RMerge2   & 11.16            & 0.89        & 124.09   & 11.1        & 1.17        & 149.18   \\
            Nsparse   & 1.49             & 0.65        & 6.64     & 1.77        & 0.84        & 7.8      \\
            AC-SpGEMM & 3.72             & 0.79        & 19.21    & 3.4         & 0.71        & 12.95    \\
            spECK     & 1.46             & 0.47        & 3.17     & 1.88        & 0.7         & 4.67     \\
            nonTCU    & 1.48             & 1.1         & 1.97     & 1.56        & 1.19        & 2.14     \\ \bottomrule
        \end{tabular}
    }
    \caption{Speedup of tSparse over the various approaches for matrices selected by our criteria.}
    \label{tab:speedup}
\end{table*}

\replaced{}{Execution time}

\replaced{}{We measure the execution time of \ore{cuSPARSE, CUSP, RMerge2, Nsparse, nonTCU and tSparse} for each matrix of our dataset \ore{in Table~}\ref{tab:data}. We examine how the characteristics of the matrices of our dataset affect the performance of each approach.
	Fig.~\ref{fig:timesTitan} shows the execution time of $A*A$ for the four approaches on the \ore{Titan RTX} GPU.
	Fig. \ref{fig:times_randomTitan} corresponds to the first part of our dataset (randomly selected matrices), whereas Fig. \ref{fig:times_criteriaTitan} corresponds to the second part (matrices selected based on criteria). Similarly, Fig. \ref{fig:times2070} shows the execution times on RTX 2070.}



\replaced{}{We make seven major observations.}

\replaced{}{First, tSparse shows better performance with denser tiles.  Grouping NZ elements in tiles reduces the number of values to work on. According to \cite{cusp_optimizing_2015}, the main cost of ESC is sorting. Consequently, fewer \replaced{values}{entries} equals to less time sorting. Generally, tSparse has good performance for \emph{bitmap density} greater than five.} 

\replaced{}{Second, tSparse generally outperforms cuSPARSE in the larger matrices of our dataset ($NNZ(A) > \sim1000000$). This happens probably because cuSPARSE does not have sufficient shared memory to store the hash tables and therefore global memory traffic increases. Nevertheless, the size of the matrix is not the only thing that decides the performance of cuSparse. To the best of our knowledge, the performance of cuSPARSE also depends on the sparsity structure of the matrix because the structure affects the number of hash conflicts.}

\replaced{}{Third, tSparse performs better than CUSP when the number of intermediate products of tiles is smaller than the number of intermediate products of elements of CUSP, i.e., $NNZ(\overline{C}\textsubscript{tiles}) < NNZ(\overline{C})$.
	Otherwise, CUSP, which handles only single elements, is faster than tSparse which has an additional overhead for handling tiles.}

\replaced{}{Fourth, tSparse is faster than Nsparse when the average row size of \emph{A} (calculated as $ NNZ(A) / Dimension $) is greater than 42. Nsparse follows a hash table approach (Section \ref{ssec:hash}). Therefore, larger rows possibly create more hash conflicts and hash tables are difficult to keep in shared memory (instead of the slower global memory) with large rows. We also note that web graph matrices (webbase-1M and web-NotreDame) have high irregularity, i.e., although the average row size is small (3.1 and 4.6 respectively), there are rows with thousands of elements (maximum 4700 and 3445 respectively). Therefore they are also taxing for Nsparse.}

\replaced{}{Fifth, RMerge2 shows behavior similar to Nsparse. tSparse has better performance than RMerge2 when the average row size of \emph{A} is greater than 42. To the best of our knowledge, RMerge2, which follows a hybrid approach (Section \ref{ssec:hybrid}), uses faster kernels when the row size is smaller than the size of a CUDA warp \cite{RMerge2_2018}.}

\replaced{}{Sixth, the execution time of all approaches follows the same trends for both GPUs, i.e., the relative execution time among matrices with the same approach and the relative execution time among approaches with the same matrix remain the same.}

\replaced{}{Seventh, Titan RTX GPU decreases the execution time for all approaches and matrices. tSparse is $1.32 \PLH $ faster on Titan RTX. cuSPARSE, CUSP, RMerge2, Nsparse, nonTCU are $1.73 \PLH $, $1.28 \PLH $, $1.51 \PLH $, $1.55 \PLH $, $1.54 \PLH $ faster, respectively.}

\replaced{}{paragraph Selection criteria
	In summary, a range of matrix characteristics define the performance of tSparse.  To help with selecting tSparse over other approaches, we define two criteria that generally work well: 1) $NNZ(A) > 1000000$, and 2) $NNZ(\overline{C}) / NNZ(\overline{C}\textsubscript{tiles}) > 9$.}

\subsection{Speedup} \label{sec:speedup}
\begin{figure*}[h!]
	\centering
	\includegraphics[width=0.75\linewidth]{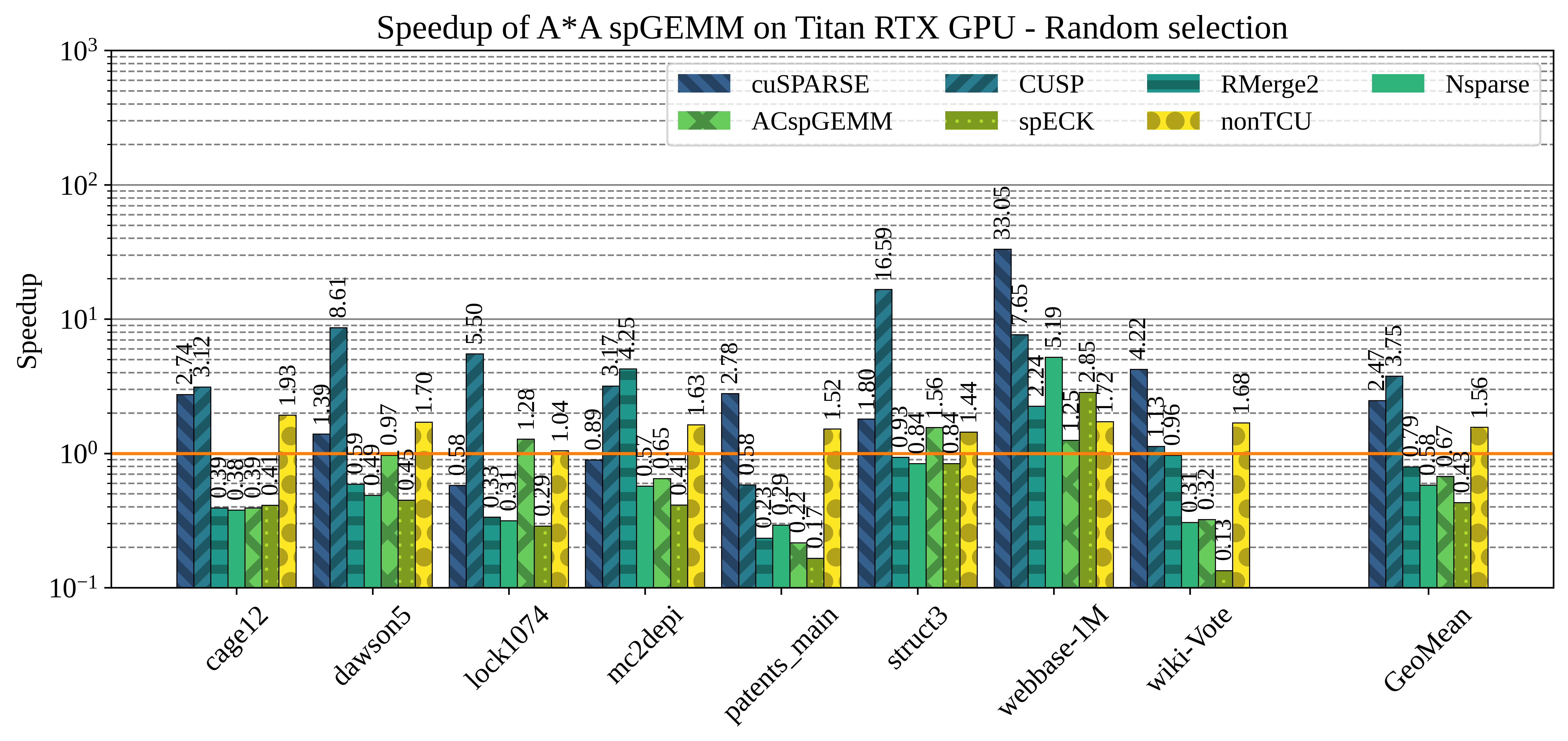}
	\caption{Speedup of tSparse on $A*A$ spGEMM using randomly selected matrices on Titan RTX.}
	\label{fig:speedup_randomTitan}
\end{figure*}

\begin{figure*}[h!]
	\centering
	\includegraphics[width=0.75\linewidth]{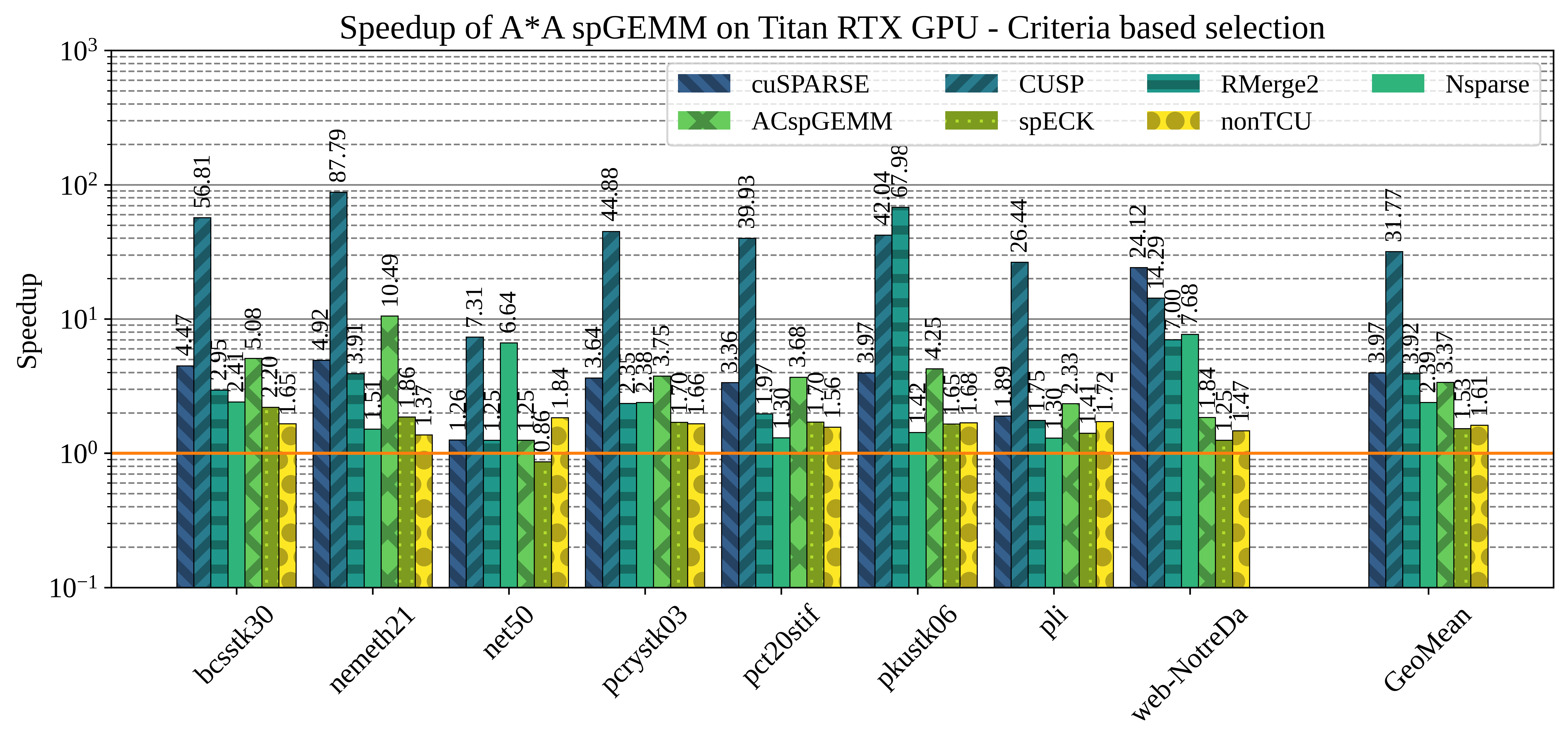}
	\caption{Speedup of tSparse on $A*A$ spGEMM using matrices selected based on criteria on Titan RTX.}
	\label{fig:speedup_criteriaTitan}
\end{figure*}

\begin{figure*}[h!]
	\centering
	\includegraphics[width=0.75\linewidth]{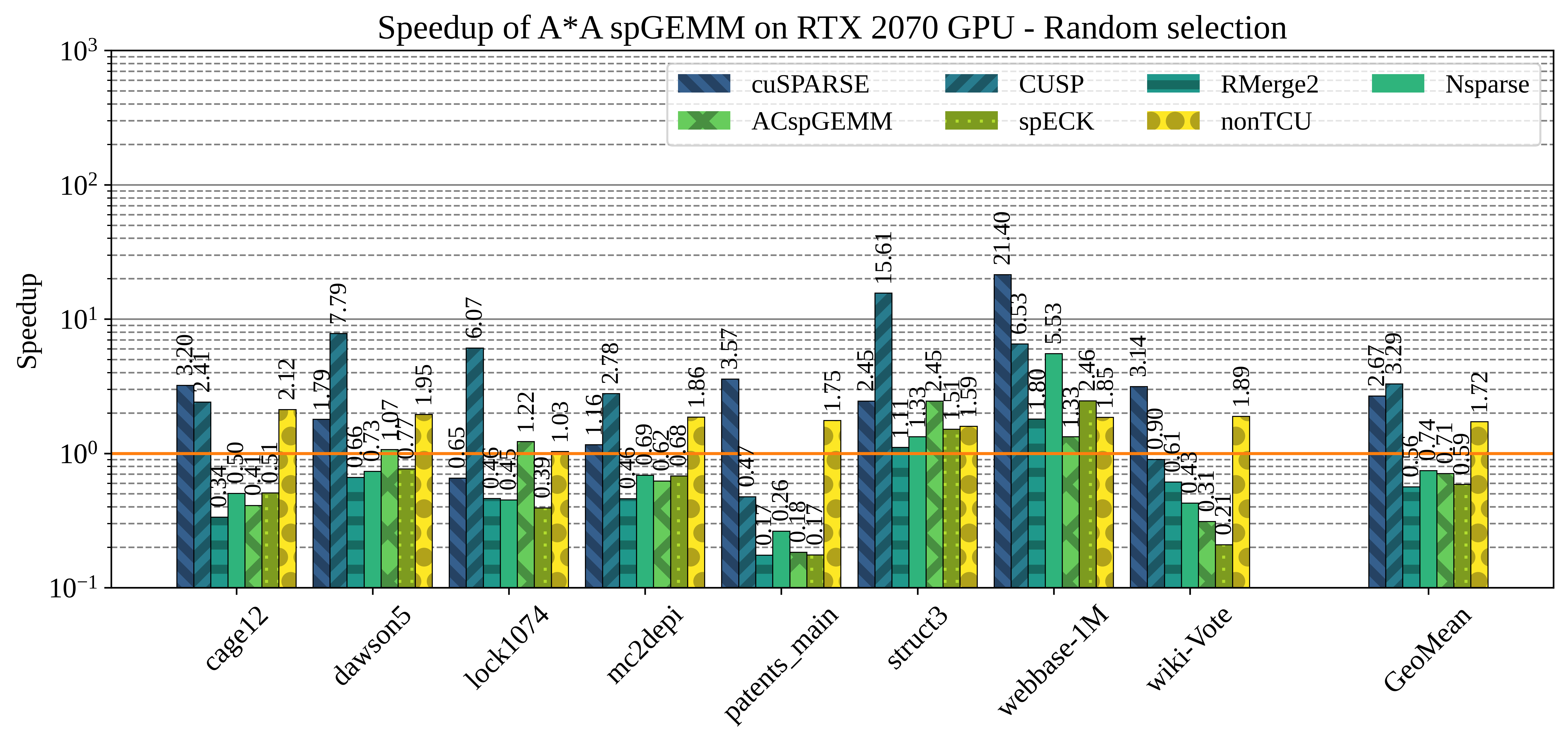}
	\caption{Speedup of tSparse on $A*A$ spGEMM using randomly selected matrices on RTX 2070.}
	\label{fig:speedup_random2070}
\end{figure*}

\begin{figure*}[h!]
	\centering
	\includegraphics[width=0.75\linewidth]{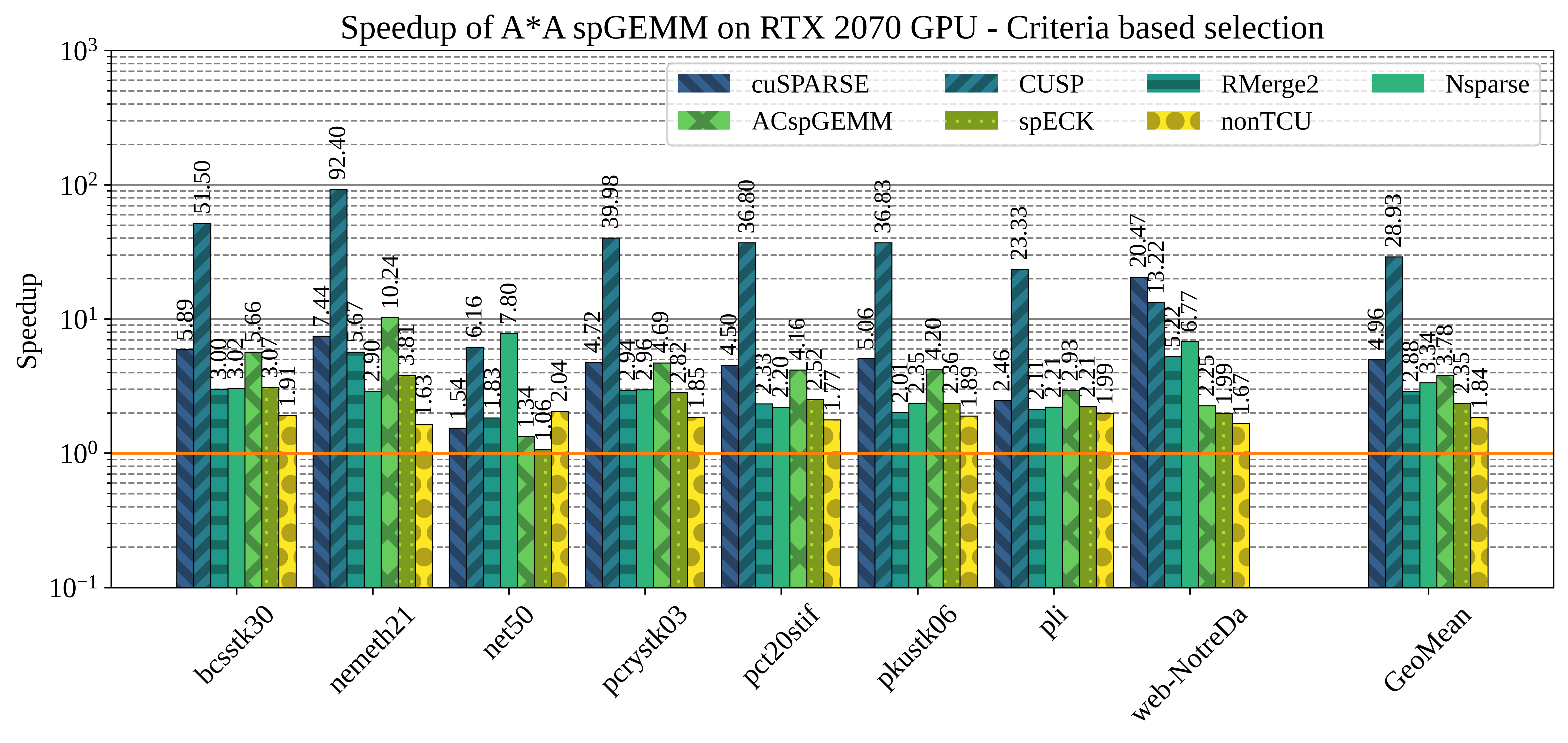}
	\caption{Speedup of tSparse on $A*A$ spGEMM using matrices selected based on criteria on RTX 2070.}
	\label{fig:speedup_criteria2070}
\end{figure*}

To show the benefits of our approach that uses TCUs for MM, we find the speedup over cuSPARSE, CUSP, RMerge2, Nsparse, AC-SpGEMM, spECK and nonTCU for \replaced{the 16}{all} matrices of our dataset.
Fig\ore{ure}~\ref{fig:speedup_randomTitan} and Fig.~\ref{fig:speedup_criteriaTitan} show the speedup of tSparse over the seven approaches when calculating $A*A$ on the \ore{Titan RTX} GPU.
Fig\ore{ure}~\ref{fig:speedup_randomTitan} corresponds to the first part of our dataset (randomly selected matrices), whereas Fig.~\ref{fig:speedup_criteriaTitan} corresponds to the second part of our dataset (matrices selected based on criteria). Figures \ref{fig:speedup_random2070} and \ref{fig:speedup_criteria2070} present the speedup on RTX 2070. GeoMean indicates the geometric average of the speedup for the matrices of the respective figure.

We make \replaced{five}{} major observations. 

First, tSparse that uses TCUs performs faster than our nonTCU implementation, by an average of $1.\replaced{68}{} \PLH$. This speedup may seem low considering that TCUs in mixed precision promise $4 \PLH$ more flops than normal fp32 operations \cite{turing_arch}. There are two reasons that keep it low: 1) our counting and multiplication kernels, \ore{without} TCUs, occupy about 50\% of the total execution time of spGEMM, so according to Amdahl's law we do not expect more than $2 \PLH$ speedup of the total execution time, and 2) tSparse is memory bound, rather than arithmetic bound, because the sorting step results in a task list with entries that point to non-continuous memory locations.

Second, tSparse outperforms cuSPARSE, CUSP, RMerge2, Nsparse, AC-SpGEMM and spECK in 14 (15), 15 (14), 10 (10), 9 (10), 11 (12) and 8 (10) out of the 16 matrices, respectively (the performance of 2070 in parentheses). The speedup of tSparse for the randomly selected matrices is: cuSPARSE $2.47 \PLH\; (2.67 \PLH)$, CUSP $3.78 \PLH\;(3.29 \PLH)$, RMerge2 $0.79 \PLH\;(0.56 \PLH)$, Nsparse $0.58 \PLH\;(0.74 \PLH)$, AC-SpGEMM $0.58 \PLH\;(0.71 \PLH)$, spECK $0.58 \PLH\;(0.59 \PLH)$, nonTCU $1.46 \PLH\;(1.72 \PLH)$.
With the matrices selected based on our criteria, the speedup of tSparse is: cuSPARSE $3.97 \PLH\;(4.96 \PLH)$, CUSP $31.77 \PLH\;(28.93 \PLH)$, RMerge2 $3.92 \PLH\;(2.88 \PLH)$, Nsparse $2.39 \PLH\;(3.34 \PLH)$, AC-SpGEMM $3.37 \PLH\;(3.78 \PLH)$, spECK $1.53 \PLH\;(2.35 \PLH)$, nonTCU $1.61 \PLH\;(1.84 \PLH)$.
The total speedup for the sixteen matrices of our dataset on average (geometric mean) is: cuSPARSE $3.12 \PLH\;(3.64 \PLH)$, CUSP $10.91 \PLH\;(9.76 \PLH)$, RMerge2 $1.76 \PLH\;(1.27 \PLH)$, Nsparse $1.18 \PLH\;(1.58 \PLH)$, AC-SpGEMM $1.51 \PLH\;(1.63 \PLH)$, spECK $0.81 \PLH\;(1.17 \PLH)$, nonTCU $1.59 \PLH\;(1.78 \PLH)$.
	
Third, tSparse owes its performance gain to both tiling and TCUs. nonTCU indicates the performance we gain by using tiling only. Across the matrices of our dataset, the speedup of nonTCU, on Titan RTX (RTX 2070), on average (geometric mean) is: cuSPARSE $1.97 \PLH\;(2.04 \PLH)$, CUSP $6.88 \PLH\;(5.49 \PLH)$, RMerge2 $1.11 \PLH\;(0.72 \PLH)$, Nsparse $0.74 \PLH\; (0.89 \PLH)$, AC-SpGEMM $0.95 \PLH\;(0.92 \PLH)$, spECK $0.51 \PLH\;(0.66 \PLH)$.

Fourth, although the spGEMM approaches generally maintain their relative ranking, the speedup of tSparse over the other approaches is greater on the RTX 2070 system. The reason is that, although the GPU parts become significantly faster on Titan RTX, the CPU tasks of tSparse (e.g., memory allocation/deallocation) take the same time on both systems. tSparse scales slightly worse than the other approaches on the higher-end GPU due to the CPU bound tasks.

Fifth, tSparse shows better performance with denser tiles.  Grouping NZ elements in tiles reduces the number of values to work on. According to \cite{cusp_optimizing_2015}, the main cost of ESC is sorting. Consequently, fewer \replaced{values}{entries} equals to less time sorting. Generally, tSparse has good performance for \emph{bitmap density} greater than five. 

\subsection{Execution time analysis} \label{sec:breakdown}
\begin{figure}[!h] 
	\centering
	\includegraphics[trim={2cm 4cm 2cm 2cm},clip,width=0.99\linewidth]{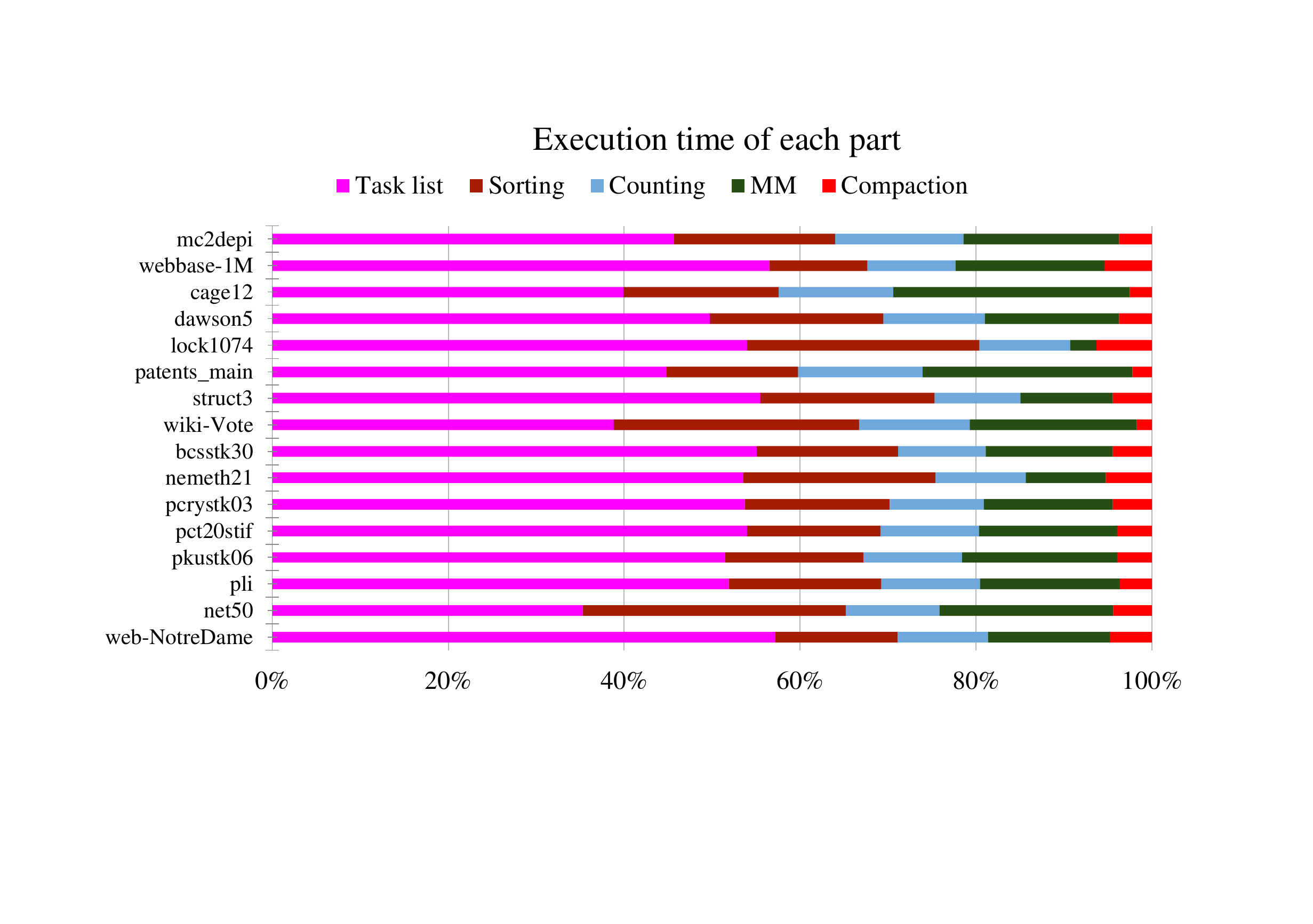}
	\caption{Relative runtime of the main parts of tSparse on Titan RTX.}
	\label{fig:breakdown}
\end{figure}

In this section, we present the relative execution time of the main parts of tSparse, i.e., Task list creation, Sorting of the task list, Counting elements of the output and the MM itself (Fig. \ref{fig:breakdown}). We make three observations. First, creating the task list takes considerable time. The main reason is that during the task list phase we do most of the necessary memory allocations. Second, Sorting takes about 19\%, Counting about 11\%, MM about 16\% and Compaction about 4\% of the execution time. Third, the runtime of Counting is not much shorter than MM's. The reason is that Counting includes the execution time for both the counting kernel and the memory allocations that the MM kernel needs.

\replaced{We note that in the execution time we include all allocations, including the allocation of the output. We also include the time for conversion from fp32 to fp16 (if the input is stored in fp32). However, we do not include the time for conversion to/from bitmap format for fair comparison to other approaches.}{}

\subsection{Numerical precision}
\begin{figure}[!h] 
	\centering
	\includegraphics[trim={0.9cm 3cm 1cm 1cm},clip,width=0.88\linewidth]{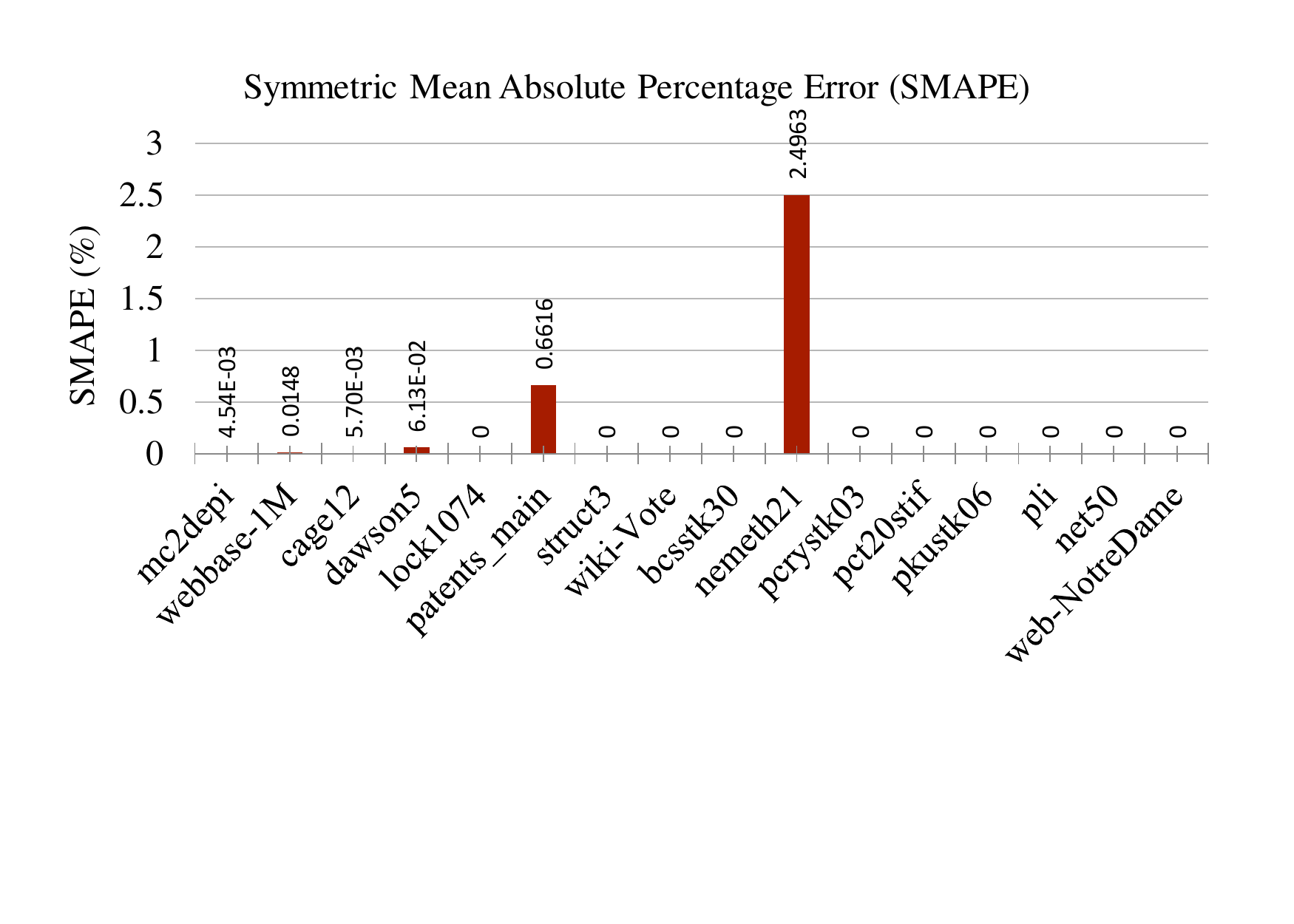}
	\caption{Numerical precision of spGEMM in Symmetric Mean Absolute Percentage Error (SMAPE).}
	\label{fig:precision}
\end{figure}

TCUs accept 16-bit numbers as input, whereas TCUs perform MM and accumulation in 32-bit precision (mixed precision - Section \ref{sec:tcus}). In this section, we show how tSparse fares w.r.t. precision in comparison to a full 32-bit approach (we use CUSP as base). Figure \ref{fig:precision} illustrates the precision of spGEMM as Symmetric Mean Absolute Percentage Error (SMAPE - Equation \eqref{eq:smape}).
\begin{equation}
	\frac{100\%}{n} \sum_{i=1}^{n}  \frac{ | x_i - \hat{x}_i| }{|x_i|+ | \hat{x}_i | } \label{eq:smape}
\end{equation}

We make two observations. First, when the input matrices contain patterns of \enquote{1} and \enquote{0} the SMAPE is 0\%. Second, when the inputs are real numbers the SMAPE is \replaced{on average}{} 0.02\%. Exception are the cases with inputs close to the limits of the fp16 range, where the round-off error is bigger (nemeth21 - 2.49\%, patents\_main - 0.66\%).

\subsection{Memory requirements}
\begin{figure}[!h] 
\centering
    \centering  
  \includegraphics[clip,  trim=0.5cm 1.0cm 0.7cm 0.1cm, width=0.99\linewidth]{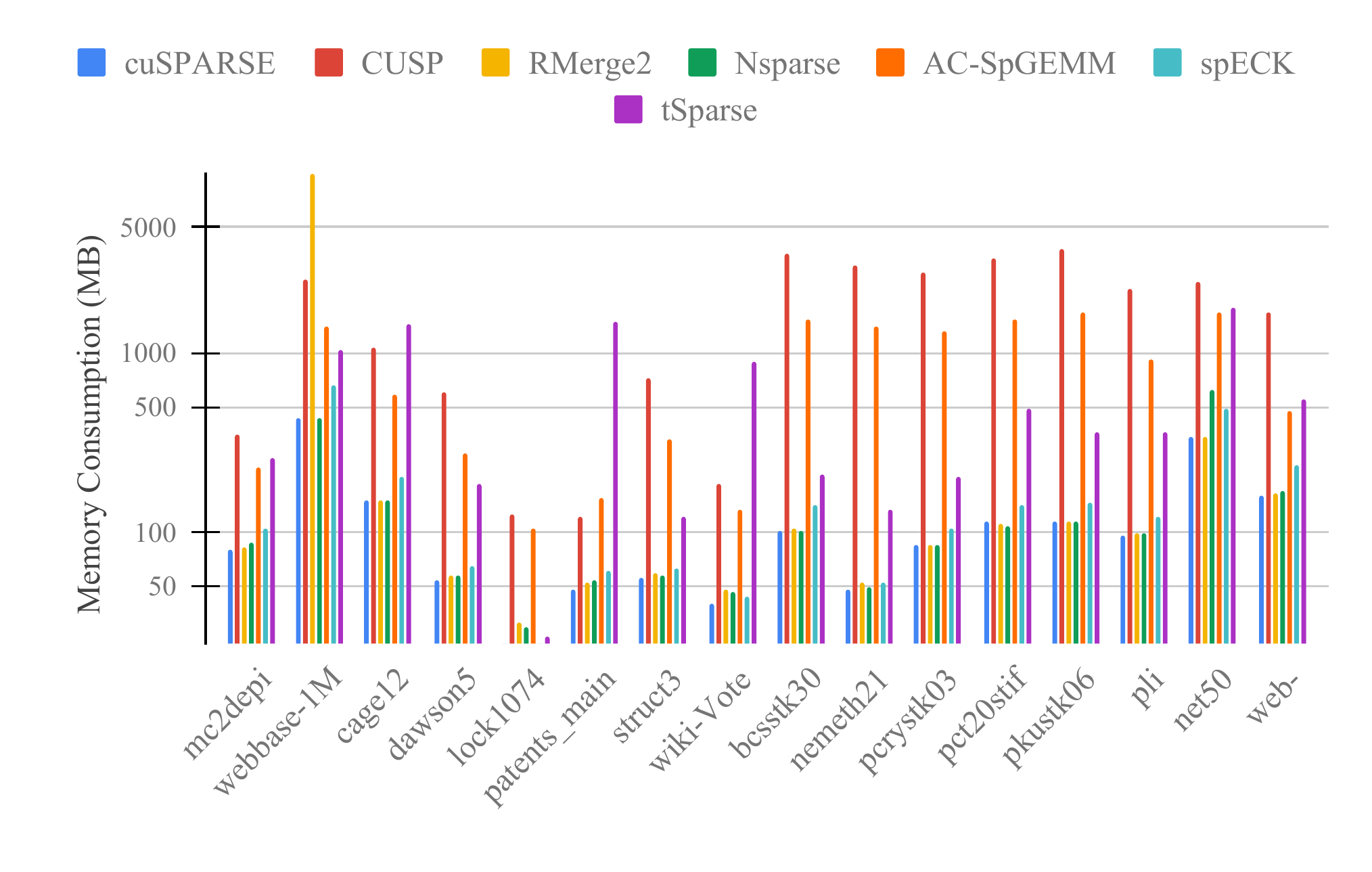}
   \caption{Memory Consumption in MB}
   \label{fig:mem_consumption}
\end{figure}

\replaced{In this section we present the peak memory consumption of all approaches. Fig. \ref{fig:mem_consumption} presents the memory consumption for the 16 matrices of our dataset.}{}

\replaced{We make four major observations. First, hash table approaches require the smallest memory area as they do not have to store huge intermediate matrices. Second, ESC approaches require large amounts of memory. Third, tSparse, which \enquote{packs} elements into tiles, typically requires more memory than hash approaches and less than ESC. The amount of memory is proportional to the bitmap density. When the average bitmap density is very low, like in patents\_main and wiki-vote, tSparse requires a lot of memory due to the additional overhead for storing tiles. Fourth, RMerge2 performs well as it usually requires only 5 bytes for each row of the left-hand side.}{}

\subsection{Summary}

We examine the performance of our approach in $A*A$ and compare to cuSPARSE, CUSP, \replaced{RMerge2, Nsparse, AC-SpGEMM and spECK}{} over the \replaced{16}{} sparse matrices of our dataset. We draw two important conclusions.
First, tSparse generally outperforms the other approaches for the larger matrices of our dataset when we have dense tiles and enough work per matrix row. Second, TCUs play an important role in the performance of our approach, speeding up nonTCU \replaced{$1.68 \PLH$}{}.

tSparse outperforms the other state-of-the-art approaches \replaced{$1.53 \PLH$ to $31.77 \PLH$}{} when our selected criteria are met, without significant loss in accuracy. Therefore tSparse is a suitable alternative for spGEMM.

\section{Related work} \label{sec:related}
To our knowledge, this is the first paper to use TCUs in the context of spGEMM. With this approach, we group NZ into tiles. Unlike previous methods, which use normal fp32 cores, we use tensor cores to multiply the tiles. We show that, with this approach, we can increase the performance of spGEMM. In this section we summarize other spGEMM implementations.



\subsection{Expansion Sorting Compression}
There exists a substantial body of work on improving ESC methodology that we describe in Section \ref{sec:esc}.
Dalton et al. \cite{cusp_optimizing_2015} improve on ESC by optimizing sorting, the most time consuming step, and by localizing processing to shared memory. Kunchum et al. \cite{kunchum_improving_2017}, in HybridSparse, implement variants of ESC. Winter et al. \cite{winter_adaptive_2019}, in AC-SpGEMM, perform ESC locally in shared memory. They use dynamic scheduling of iterations of ESC to keep data longer in shared memory. Thus, they reduce global memory traffic and the cost of sorting a huge intermediate matrix in global memory.

tSparse moves expansion and compression steps after sorting. This way we reduce memory allocation and movements and utilize MAC of TCUs.


\subsection{Hash tables} \label{ssec:hash}
Hash tables can mitigate the cost of sorting and storing huge intermediate matrices.
Demouth et al. \cite{cusparse_presentation_2012} present one of the first implementations of spGEMM with hash tables. cuSPARSE \cite{CUDAProgrGuide} is based on the work of Demouth et al. Their approach has two drawbacks. First, there is imbalance between threads because different threads of the warp \replaced{might}{my} have to insert a different number of values in the hash table. Second, shared memory space is limited, which results in frequent data movement to global memory. Anh et al. \cite{anh_balanced_2016}, in BalancedHash,  and Nagasaka et al. \cite{nagasaka_high-performance_2017}, in Nsparse, reduce the consumption of shared memory by partitioning the rows of the input or output, respectively. Nsparse improves on BalancedHash in two ways: 1) by using hash tables of variable size in shared memory, less shared memory is required and more thread blocks can run, and 2) by using fewer auxiliary matrices, it keeps memory traffic low and reduces memory storage requirements. Deveci et al. \cite{deveci2018multithreaded} use two-level hash tables. They adapt the hash tables to the number of threads in order to create a method that is portable to many platforms. \replaced{Parger et al. \cite{parger_speck_2020}, in spECK find a trade-off between analysis cost for load balancing and expected gain.}{}

In tSparse, 1) we have less values to sort because elements are grouped into tiles, and 2) we avoid storing the intermediate products by using the task list to directly accumulate them.


\subsection{Hybrid} \label{ssec:hybrid}
Hybrid methods select among multiple methods/kernels during spGEMM depending on the workload.
Dalton et al. \cite{cusp_optimizing_2015} improve on ESC methodology by changing the thread granularity of the sorting method based on the size of rows of \emph{C}. Liu et al. \cite{bhsparse_2015}, in bhSparse,  change the sorting-merging method depending on the size of rows of \emph{C}. Kunchum et al. \cite{kunchum_improving_2017}, in HybridSparse, choose the spGEMM method based on the workload of each row of \emph{A}. They select among variants of ESC and their own method (a GPU implementation of scatter vectors). Gremse et al. \cite{RMerge_2015,RMerge2_2018}, in RMerge create different kernels for different row sizes of \emph{A}.
Hybrid methods achieve good load balance thanks to their adaptability to the workload.
 
We can divide tSparse in two parts. The part that performs ESC on tiles to form the task list and the part that uses our kernels to perform MAC operations. For the first part, we use Thrust which achieves good load balance. For the second part, we let the GPU hardware scheduler manage the workload. This is possible because we employ many thread blocks, which the GPU schedules to SMs based on the available resources of each SM.

\section{Conclusion} \label{sec:conclusion}
In this work, we utilize TCUs to increase the performance of spGEMM. To that end, we modify the ESC method and we create a task list of MMs of tiles. 
The key advantages of our approach, tSparse, are two. First, tiles reduce the number of values that the computationally demanding parts of ESC have to act on. Second, the task list sends the tiles to TCUs, which not only perform MM faster than normal computation cores, but also leave the normal cores free for different workloads.

The results confirm that TCUs increase the performance of MM and the combination of our tiling approach with TCUs provides significant benefits to spGEMM. TCUs increase the performance of tSparse by \replaced{68}{60}\% in comparison to our nonTCU implementation. Our approach is, on average, \replaced{$1.53 \PLH$ to $31.77 \PLH$}{} faster than cuSPARSE, CUSP, RMerge2, Nsparse, \replaced{AC-SpGEMM and spECK when $NNZ(A) > 300000$ and $\widehat{RowA} > 42$}{}. We conclude that our methodology improves the performance of spGEMM by making efficient use of tiles and TCUs. The source code of our approach is available at \url{https://github.com/oresths/tSparse}.

%

	\section*{Acknowledgment}
This work is supported by High Performance Soft-tissue Navigation ( HIPERNAV - H2020-MSCA-ITN-2016 ). HIPERNAV has received funding from the European Union’s Horizon 2020 research and innovation programme under grant agreement No 722068.


\bibliographystyle{elsarticle-num}
\bibliography{spmm}


\vspace{10pt}
\setlength{\intextsep}{0pt}%
\noindent \textbf{Orestis Zachariadis} is a researcher and PhD candidate at University of Cordoba. He works on  parallel computing for medical image registration algorithms as part of HiPerNav EU project. In 2018, he worked on medical technology at SINTEF. In 2016, he joined MultiDrone EU project as a research assistant at University of Bristol, where he worked on object and human detection and following in videos shot from multiple drones. He holds a BSc and an MSc in Electrical and Computer Engineering from Aristotle University of Thessaloniki. His research interests include parallel computing, computer vision and robotics.\par

\vspace{10pt}
\setlength{\intextsep}{0pt}%
\noindent \textbf{Nitin Satpute} is a researcher and PhD candidate at University of Cordoba. He works on GPU acceleration for medical image enhancement and segmentation algorithms as a part of HiPerNav EU project. In 2018, he worked on liver segmentation at OUH, Oslo, Norway. In 2019, he worked on liver enhancement at NTNU Gjovik Norway, where he developed fast parallel cross modality approach for liver contrast enhancement. He holds Master of Engineering in Embedded Systems from BITS Pilani, India. His research interests include GPU computing, Medical Imaging, Re-configurable Computing and Video Processing.\par

\vspace{10pt}
\noindent \textbf{Juan G\'omez Luna} is a postdoctoral researcher at ETH Z\"urich. He received the BS and MS degrees in Telecommunication Engineering from the University of Sevilla, Spain, in 2001, and the PhD degree in Computer Science from the University of Córdoba, Spain, in 2012. Between 2005 and 2017, he was a lecturer at the University of C\'ordoba. His research interests focus on Processing-in-Memory, GPUs and heterogeneous systems, medical imaging, and bioinformatics.\par

\vspace{10pt}
\noindent \textbf{Joaqu\'in Olivares} was born in Elche, Spain. He received the B.S. and M.S. degree in Computer Sciences in 1997, and 1999, respectively, and the M.S. degree in Electronics Engineering in 2003, all from the Universidad de Granada, Spain. He received the Ph.D. degree in 2008 at the Universidad de Cordoba, Spain. He is Associate Professor with the Electronic and Computer Engineering Department at the Universidad de Cordoba, Spain, since 2001. He is founder and head of the Advanced Informatics Research Group. His research interests are in the field of embedded systems, computer vision, and high performance computing.\par

%

\end{document}